\documentclass[aps,showpacs,preprintnumbers,amsmath,amssymb,nofootinbib,eqsecnum,onecolumn,preprintnumbers]{revtex4}
\usepackage{graphicx}
\usepackage{epsf}
\usepackage{amsmath}
\usepackage{epstopdf}
\usepackage{bm}
\usepackage{color}
\usepackage{tabularx}
\usepackage{enumitem}
\usepackage{float}
\usepackage{array,booktabs}
\usepackage{footnote}
\usepackage{threeparttable}
\usepackage{graphicx}
\usepackage{hyperref}
\usepackage{amssymb,epsf}
\usepackage{latexsym}
\usepackage{epstopdf}
\usepackage{epsfig}

\begin{document}
\title{Charged black hole chemistry with massive gravitons}
\author{Ali Dehghani$^1$\footnote{email address: ali.dehghani.phys@gmail.com},
Seyed Hossein Hendi$^{1,2}$\footnote{email address: hendi@shirazu.ac.ir}} \affiliation{$^1$
Physics Department and Biruni Observatory, College of Sciences,
Shiraz University, Shiraz 71454, Iran \\
$^2$ Research Institute for Astrophysics and Astronomy of Maragha
(RIAAM), P.O. Box 55134-441, Maragha, Iran}

\begin{abstract}
In the subject of black hole chemistry, a broad variety of
critical phenomena for charged topological black holes (TBHs) with
massive gravitons (within the framework of dRGT massive gravity)
is discussed in detail. Since critical behavior and nature of
possible phase transition(s) crucially depend on the specific
choice of ensemble, and, in order to gain more insight into
criticality in the massive gravity framework, we perform our
analysis in both the canonical (fixed charge, $Q$) and the grand
canonical (fixed potential, $\Phi$) ensembles. It is shown that,
for charged TBHs in the grand canonical ensemble, the van der
Waals (vdW) phase transition could take place in $d \ge 5$, the
reentrant phase transition (RPT) in $d \ge 6$ and the analogue of
triple point in $d \ge 7$ which are different from the results of
canonical ensemble. In the canonical ensemble, the vdW phase
transition is observed in $d \ge 4$, the vdW type phase transition
in $d \ge 6$ and the critical behavior associated with the triple
point in $d \ge 6$. In this regard, the appearance of grand
canonical $P-V$ criticality and the associated phase transition(s)
in black holes with various topologies depend on the effective
topological factor $k_{\rm{eff}}^{\rm{(GC)}} \equiv {[k +
{m_g^2}c_0^2{c_2} - 2({d_3}/{d_2}){\Phi ^2}]}$ instead of $k$ in
Einstein gravity, where $k$ is the normalized topological factor
($k_{\rm{eff}}^{\rm{(C)}} \equiv [k + {m_g^2}c_0^2{c_2}]$ plays
this role in the canonical ensemble of TBHs in massive gravity).
Such evidence gives the (grand) canonical study of extended phase
space thermodynamics with massive gravitons a special
significance.

\end{abstract}

 \maketitle

\section{Introduction}
Asymptotically anti-de Sitter (AdS) black holes admit certain
phase transitions in Einstein gravity, e.g., Hawking-Page phase
transition (a transition between the AdS spacetime and the AdS
black hole) \cite{HawkingPage1983} is observed for a variety of
asymptotically AdS black hole configurations such as
Schwarzschild-AdS, Reissner-Nordstr\"{o}m-AdS (RN-AdS) and Kerr
(-Newman) AdS ones \cite{HawkingPage1983,EmparanChamblin1999a,
EmparanChamblin1999b, Caldarelli2000KerrNewmanAdS}.  Seeking for a
transition from a black hole phase to another phase, the van der
Waals (vdW) behavior has been found for charged-AdS
\cite{EmparanChamblin1999a,EmparanChamblin1999b,KubiznakMann2012}
and Kerr (-Newman) AdS black holes
\cite{Caldarelli2000KerrNewmanAdS,Altamirano2012Reentrant,Wei2016KerrNewmanAdS}.
Further investigations on the phase transition between black hole
phases have been revealed that the reentrant phase transition
(RPT) present in liquid crystals and multicomponent fluids
\cite{NarayananKumar1994} can occur for Kerr-AdS black holes in
$d\geq6$ dimensions \cite{Altamirano2012Reentrant}. In addition,
the small/intermediate/large black hole (SBH/IBH/LBH) phase
transition associated with the triple point was found for
multi-spinning Kerr-AdS black holes ( in $d\geq6$)
\cite{Altamirano2014TriplePoint} which is interpreted as
solid/liquid/gas phase transition typical of many materials. These
critical behaviors have been observed in a vast range of black
hole configurations including nontrivial electromagnetic fields
\cite{Fernando(2006)BI-AdS,Gunasekaran2012Mann,Vahidinia2013} and
higher order curvatures
\cite{Cai2013GaussBonnet,Zou2014GB-PV-GrandCan,PV2014Lovelock,PV2014LovelockBI-Mo,Frassino2014,PV2015LovelockBI-Belhaj,PV2015Lovelock-nonExtended,HennigarBrennaMann2015,LovelockRainbow2017,Hennigar2017BlackBranesPV,LovelockHairyBHs2017,Mir2019Mann}.
But the story is different in dRGT massive gravity theory
\cite{dRG2010,dRGT2011} since all topological black holes (TBHs)
can experience critical behavior and phase transition due to the
massive graviton self-interaction potentials
\cite{Hendi2017Mann-PRD,Cai2015massive,PVmassive2015PRD,DehghaniHendiMann2019,Reentrant-dRGTmassive-2017,Triple-BI-massive-2017,HendiDehghani2019}.
From the AdS/CFT perspective \cite{Witten1998a,Witten1998b}, it
leads to new prospects for investigating critical behavior of TBHs
since there exists a dual interpretation in the gauge field theory
(CFT) living on the AdS boundary.

These analogies between the standard thermodynamic phase
transitions and the black hole ones have been found in the
extended phase space (varying cosmological constant as pressure,
$\Lambda = -8 \pi P$ \cite{Kastor2009CQG})
\cite{KubiznakMann2012,Altamirano2012Reentrant,Altamirano2014TriplePoint,Gunasekaran2012Mann,Vahidinia2013,Cai2013GaussBonnet,Zou2014GB-PV-GrandCan,PV2014LovelockBI-Mo,Frassino2014,PV2015LovelockBI-Belhaj,HennigarBrennaMann2015,LovelockRainbow2017,Hennigar2017BlackBranesPV,LovelockHairyBHs2017,Mir2019Mann,Hendi2017Mann-PRD,PVmassive2015PRD,DehghaniHendiMann2019,Reentrant-dRGTmassive-2017,Triple-BI-massive-2017,HendiDehghani2019}
and in some cases in the non-extended phase space (fixed
cosmological constant)
\cite{EmparanChamblin1999a,EmparanChamblin1999b,Caldarelli2000KerrNewmanAdS,Fernando(2006)BI-AdS,PV2015Lovelock-nonExtended,Cai2015massive,Dehyadegari2019}.
For example, Reissner-Nordstr\"{o}m-AdS black holes possess a
first order phase transition with swallowtail behavior which
closely resembles the well-known vdW phase transition in fluids in
both non-extended \cite{EmparanChamblin1999a} and extended
\cite{KubiznakMann2012} phase spaces with the same critical
exponents as the vdW system. But, these analogies in the
non-extended phase space are confusing since some black hole's
intensive (extensive) quantities have to be identified with
irrelevant extensive (intensive) quantities in the fluid system;
for example, the identification between the fluid temperature and
the $U(1)$ charge of RN-AdS black holes
\cite{EmparanChamblin1999a,EmparanChamblin1999b} is really
puzzling. By employing the extended phase space, these kinds of
mismatches will be eliminated \cite{KubiznakMann2012}. Motivated
by this fact, the extended phase space thermodynamics (first
established in \cite{Kastor2009CQG} and then developed in
\cite{Dolan2011CQG1,Dolan2011CQG2,Dolan2011PRD,CveticKubiznak2011Gibbons-PRD,DolanKastorMann2013})
is of direct interest for the present day, and, from this
perspective, black holes can be understood from the viewpoint of
standard chemistry, known as Black Hole Chemistry
\cite{BlackHoleChemistry2017CQG}.

In Einstein gravity, charged or rotating AdS black holes only
admit phase transition in canonical (fixed $Q$ or fixed $J$)
ensemble \cite{KubiznakMann2012,Galaxies2014} whereas, as we will
see in this paper, this statement is no longer valid in massive
gravity. However, for spherically symmetric black holes in
Einstein gravity coupled with nonlinear electromagnetic sources
such as power Maxwell invariant (PMI) electrodynamics, $P-V$
criticality is observed in both canonical (fixed $U(1)$ charge)
and grand canonical (fixed $U(1)$ potential) ensembles of the
extended phase space \cite{Vahidinia2013}. Using the canonical and
grand canonical analysis, it was shown that the non-extended phase
space of the charged-AdS black holes in dRGT massive gravity
accepts first order phase transitions in a way reminiscent of vdW
systems \cite{Cai2015massive}. Moreover, AdS black hole solutions
within the framework of higher curvature gravities exhibit rich
(non) extended phase space with the associated critical behaviors
in both the canonical and the grand canonical ensembles
\cite{Cai2013GaussBonnet,Zou2014GB-PV-GrandCan,PV2014Lovelock,PV2014LovelockBI-Mo,Frassino2014,PV2015LovelockBI-Belhaj,HennigarBrennaMann2015,LovelockRainbow2017,Hennigar2017BlackBranesPV,HennigarMann2017lambdaLine,HendiDehghani2019}.
Therefore, it seems geometrical modifications of general
relativity (such as massive gravity, Lovelock gravity, $F(R)$
gravity, ...) or the presence of nontrivial energy-momentum
tensors (such as PMI or Born-Infeld electrodynamics) are mandatory
to have black hole phase transitions in the grand canonical
ensemble where electric potential $\Phi$ or angular velocity
$\Omega$ are fixed.

One of the geometrical modifications of General Relativity (GR) is the dRGT massive gravity which is regarded as a consistent extension of Einstein's GR with an explicit mass term for spin-2 gravitons \cite{dRG2010,dRGT2011}. This alternative theory of gravity modifies GR in the large scales (IR limit) and has
some nice properties such as being ghost free
\cite{dRGT2011,HassanRosen2012PRL}, in agreement with recent observational data of LIGO collaboration
\cite{LIGO2017,deRham2014Review}, and the ability to explain the current observations related to dark matter
\cite{Schmidt-May2016DarkMatter} and also the accelerating
expansion of universe without requiring any dark energy component
\cite{MassiveCosmology2013,MassiveCosmology2015}. On the other hand, the possibility of embedding massive gravity in an
ultraviolet-complete theory like string theory is indicated in
Ref. \cite{MGinString2018}. In this paper, the authors considered
4-dimensional AdS solutions of IIB string theory in which the
lowest spin-2 mode has a tiny mass ($m_{g}$), and explicitly
showed that massive AdS$_{4}$ gravity is a part of the string
theory landscape \cite{MGinString2018}. For these reasons we
regard massive gravity as a class of theories that merit further
exploration.

Massive gravity theories need an auxiliary reference metric
(${f_{\mu \nu }}$) to define a mass term for massless gravitons
\cite{deRham2014Review,Hinterbichler2012Review}, and, in
principle, one can construct a special theory of massive gravity
for each choice of reference metric \cite{HassanRosen2012JHEP}.
Massive gravity theories in AdS space with a singular (degenerate)
reference metric have been particularly useful in the context of
gauge/gravity duality, where a finite DC conductivity obtains when
studying the dual boundary theory
\cite{Vegh2013,BlakeTong2013,Davison2013}, in contrast to massless
gravity theories such as Einstein and Gauss-Bonnet ones with
infinite DC conductivity
\cite{HHH2008PRL,HHH2008JHEP,Gregory2009GBsuperconductor,Gregory2010GBsuperconductor}. On the other hand,
AdS black hole solutions in massive gravity theories can
effectively describe different phases of condensed matter systems
with broken translational symmetry such as solids, liquids,
(perfect) fluids etc
\cite{Alberte2015,Alberte2016SolidHolography,Alberte2018HolographicPhonons,Alberte2018BHelasticity}.
As reported in Ref. \cite{DehghaniHendiMann2019}, one can build a
gravitational theory dual system with this property and then see
if it can thermodynamically simulate the critical behavior of
those condensed matter systems as well. To do that, it is shown
that in parallel with everyday thermodynamics, one has to insert a
$P-V$ term for the AdS black hole systems and this can be done in
the context of BHC \cite{BlackHoleChemistry2017CQG} by extending
the thermodynamic phase space \cite{Kastor2009CQG}. One of the
interesting cases is the chemistry of charged TBHs in massive
gravity, since these black objects can be viewed as effective dual
field theory of different types of charged condensed matter
systems. Hence in this paper, we study charged BHC with massive
gravitons with three reasonable reasons/goals: i) Considering TBHs
in a more complicated environment via modified gravity, here dRGT
massive gravity, leads to more possibilities for investigation.
ii) In the context of BHC via dRGT massive gravity, since
holographic phase transitions take place in both the canonical and
grand canonical ensembles (as will be shown here), we can learn
more about the effect of ensemble one is dealing with and keep
track the outcomes of critical phenomena in each ensemble,
separately. iii) We intend to generalize the results of neutral
massive TBHs to the charged cases and see what happens when a
$U(1)$ charge is added. On the other hand, the inclusion of a
$U(1)$ charge is equivalent to state that the charged TBHs may
resemble those critical behaviors present in charged condensed
matter systems.

So far, all studies related to the extended phase space
thermodynamics and critical behaviors (only for vdW phase
transition) of charged TBHs in massive gravity have been performed
in the canonical ensemble. In addition, for the other types of
phase transitions such as the RPT or triple point phenomena, only
the uncharged (neutral) TBHs have been studied
\cite{Reentrant-dRGTmassive-2017,DehghaniHendiMann2019}. In this
regard, the RPT and triple point phenomena are not yet observed in
the charged black holes of massive gravity. Here, we explicitly
consider the effects of $U(1)$ charge and potential to investigate
the associated $P-V$ criticality and bring out several important
results. Since the specific choice of ensemble has an important
role on the criticality and nature of possible phase
transition(s), it is of great interest to generalize these studies
in the extended phase space of TBHs to the grand canonical
ensemble. In fact, as we will explicitly show that a certain
critical behavior in a specific $d$-dimensional spacetime could be
present or absent depending on the nature of the ensemble one is
dealing with.

Taking the above mentioned motivations into account, we intend to
develop the extended phase space of the dRGT black hole solutions
in the grand canonical ensemble along with the canonical ensemble
by applying suitable boundary conditions. Our purpose is to
discover which properties of AdS black holes are universal and
which ones show a dependence on the spacetime dimensions and the
ensemble. To do that, this paper is structured as follows: first,
in Sect. \ref{Bulk theory}, considering the full nonlinear theory
of dRGT massive gravity in higher dimensions, we present exact
charged black hole solutions with appropriate boundary conditions.
Then, in Sect. \ref{Canonical BHC}, the subject of charged black
hole chemistry in massive gravity is investigated in the canonical
ensemble. Afterward, in Sect. \ref{Grand Can. BHC}, the context of
charged black hole chemistry in massive gravity is promoted to the
grand canonical ensemble, and then, in Sect. \ref{critical
exponents}, we compute the associated critical exponents in both
the canonical and grand canonical ensembles. Finally, in Sect.
\ref{Conclusion}, we finish our paper with some concluding remarks
and summarize the results.

\section{Action, field equations, and topological black holes} \label{Bulk theory}
The total action, ${{\cal I}_G}$, for a gravitating system consists of three terms as
\begin{equation}
{{\cal I}_G} = {{\cal I}_b} + {{\cal I}_s} + {{\cal I}_{ct}},
\end{equation}
where ${\cal I}_b$, ${{\cal I}_s}$ and ${\cal I}_{ct}$ are called
the bulk action, the surface term (boundary action), and the
counterterm action, respectively. The bulk action for dRGT massive
gravity on the $d = (n+2)$-dimensional background manifold $\cal
M$ in the presence of negative cosmological constant ($\Lambda =
-\frac{{d_1}{d_2}}{2 \ell ^2}$, with the AdS radius $\ell$) and
Maxwell invariance $\cal F$ is
\begin{eqnarray} \label{bulk action}
{{\cal I}_{\rm{b}}} =  - \frac{1}{{16\pi {G_d}}}\int_{\cal M} {d^d}x\sqrt { - g} \,[R - 2\Lambda  - {\cal F} +
 m_g^2\sum\limits_{i=1}^{d-2} {{c_i}{{\cal U}_i}(g,f)}],
\end{eqnarray}
where ${\cal F} \equiv {F^{\mu \nu }}{F_{\mu \nu }}$, and $F_{\mu
\nu }$ is the Faraday tensor which is constructed using the $U(1)$
gauge field ${A_\nu }$ as ${F_{\mu \nu }} = {\partial _{[\mu
}}{A_{\nu ]}}$. In the above action, $m_g$ is the graviton mass
parameter, $c_{i}$'s are massive couplings which are arbitrary
constants, ${g_{\mu \nu }}$ is the physical metric, and ${f_{\mu
\nu }}$ is a fixed second rank symmetric tensor as an auxiliary
reference metric to define a mass term for massless spin-2
particles (i.e., gravitons). As shown in Ref.
\cite{dRG2010,dRGT2011}, gravitons become massive without any
ghost problem \cite{HassanRosen2012PRL} if one adds the
interaction potentials ${{\cal U}_i}$ to the Lagrangian density of
Einstein's gravity \footnote{This statement holds in higher
dimensions as well. The higher dimensional extension of the
massive (bi)gravity, including higher order graviton's
self-interactions, is discussed in Ref. \cite{TQDo2016a,TQDo2016b}
which confirms the absence of ghost fields using the
Cayley-Hamilton theorem.}. The graviton interaction terms are
symmetric polynomials of the eigenvalues of $d \times d$ matrix
${\cal K}_{\,\,\,\,\,\nu }^\mu  = \sqrt {{g^{\mu \lambda
}}{f_{\lambda \nu }}}$ and may be written as \footnote{Our
notation is a little different with Refs.
\cite{dRG2010,dRGT2011,HassanRosen2012PRL} and is in agreement
with Refs. \cite{Vegh2013,Hendi2016JHEP,Hendi2017Mann-PRD}}
\begin{equation}
{{\cal U}_i} = \sum\limits_{y = 1}^i {{{( - 1)}^{y + 1}}\frac{{(i - 1)!}}{{(i - y)!}}} {{\cal U}_{i - y}}[{{\cal K}^y}],
\end{equation}
in which the square root of $\cal K$ stands for matrix square root
(i.e., $(\sqrt {\cal K} )_{\,\,\,\,\lambda }^\mu (\sqrt {\cal K}
)_{\,\,\,\,\nu }^\lambda  = {\cal K}_{\,\,\,\,\,\nu }^\mu $) and
the rectangular brackets denote traces, $[{\cal K}] = {\cal
K}_{\,\,\,\,\mu }^\mu$. Some explicit form of ${\cal U}_{i}$'s are
given as
\begin{eqnarray}
{{\cal U}_1} &=& [{\cal K}], \nonumber\\
{{\cal U}_2}& = &{[{\cal K}]^2} - [{{\cal K}^2}], \nonumber\\
{{\cal U}_3}& = &{[{\cal K}]^3} - 3\,[{\cal K}]\,[{{\cal K}^2}] + 2\,[{{\cal K}^3}], \nonumber\\
{{\cal U}_4} &= &{[{\cal K}]^4} - 6\,{[{\cal K}]^2}\,[{{\cal K}^2}] + 8\,[{\cal K}]\,[{{\cal K}^3}] + 3\,{[{{\cal K}^2}]^2} - 6\,[{{\cal K}^4}], \nonumber\\
{{\cal U}_5}& =& {[{\cal K}]^5} - 10\,[{\cal K}]{\,^3}[{{\cal K}^2}] + 20{[{\cal K}]^2}\,[{{\cal K}^3}] - 20[{{\cal K}^2}]\,[{{\cal K}^3}] \nonumber\\
&&+ 15\,[{\cal K}]{[{{\cal K}^2}]^2} - 30\,[{\cal K}]\,[{{\cal K}^4}] + 24\,[{{\cal K}^5}], \nonumber\\
\, \vdots
\end{eqnarray}

Using the variational principle, the electromagnetic and
gravitational field equations of the bulk action (\ref{bulk
action}) can be obtained. Varying the bulk action with respect to
the dynamical metric (${g_{\mu \nu }}$) and gauge field
(${A_{\nu}}$) results
\begin{eqnarray} \label{bulk variation}
\delta {{\cal I}_{\rm{b}}} &=&  - \frac{1}{{16\pi {G_d}}}\int_{\cal M} {{d^d}x\sqrt { - g}} [{G_{\mu \nu }} + \Lambda {g_{\mu \nu }} + m_g^2{{\cal X}_{\mu \nu }} - {T_{\mu \nu }}] \delta {g^{\mu \nu }}\nonumber\\
&&+ \frac{1}{{8\pi {G_d}}}\int_{\partial {\cal M}} {{d^{d - 1}}x\sqrt { - h} } {n^\alpha }{h^{\mu \nu }}\delta {g_{\mu \nu ,\alpha }} \nonumber\\
&&- \frac{1}{{4\pi {G_d}}}\int_{\cal M} {{d^d}x\sqrt { - g} } [{\nabla _\mu }{F^{\mu \nu }}]\delta {A_\nu }\nonumber\\
&&+ \frac{1}{{4\pi {G_d}}}\int_{\partial {\cal M}} {{d^{d - 1}}x\sqrt { - h} } {n_\mu }{F^{\mu \nu }} \delta {A_\nu },
\end{eqnarray}
where $n_{\mu}$ is a radial unit vector pointing outwards and
${h_{\mu \nu}}$ is the induced metric of the boundary (${\partial
{\cal M}}$) \footnote{It should be noted that, in the massless
limit of massive gravity, some problematic boundary terms induced
by the bulk action appear and one should introduce novel boundary
counterterms which dominate over the Gibbons-Hawking term and
cancel those terms \cite{Gabadadze2018Pirtskhalava}. In the
present work, considering the massive graviton case, no further
boundary term is needed since all the new fields of massive
gravity enter the action with the first derivative, so do not
alter the equations of motion.}. In the above expression, ${T_{\mu
\nu }}$ and ${\cal X}_{\mu \nu }$ are the consequences of varying
the Maxwell invariant and graviton interaction potentials with
respect to the ${g_{\mu \nu }}$ as below
\begin{equation}
{T_{\mu \nu }} =  - \frac{1}{2}{g_{\mu \nu }}{\cal F} + 2{F_{\mu \lambda }}{F_\nu }^{\,\lambda },
\end{equation}
\begin{equation} \label{massive filed terms}
{{{\cal X}_{\mu \nu }} =  - \sum\limits_{i=1}^{d-2} {\frac{{{c_i}}}{2}\Big[{{\cal U}_i}{g_{\mu \nu }} + \sum\limits_{y = 1}^i {{{( - 1)}^y}\frac{{i!}}{{(i - y)!}}{{\cal U}_{i - y}}{\cal K}_{\mu \nu }^y} \Big]} },
\end{equation}
where, in our notation, ${{\cal U}_{i - y}} = 1$ if $i=y$. The explicit form of ${\cal X}_{\mu \nu }$ can be presented as
\begin{eqnarray}
{{\cal X}_{\mu \nu }} &=&  - \frac{{{c_1}}}{2}({{\cal U}_1}{g_{\mu \nu }} - {{\cal K}_{\mu \nu }}) \nonumber\\
&&- \frac{{{c_2}}}{2}({{\cal U}_2}{g_{\mu \nu }} - 2{{\cal U}_1}{{\cal K}_{\mu \nu }} + 2{\cal K}_{\mu \nu }^2) \nonumber\\
&& - \frac{{{c_3}}}{2}({{\cal U}_3}{g_{\mu \nu }} - 3{{\cal U}_2}{{\cal K}_{\mu \nu }} + 6{{\cal U}_1}{\cal K}_{\mu \nu }^2 - 6{\cal K}_{\mu \nu }^3) \nonumber\\
&&- \frac{{{c_4}}}{2}({{\cal U}_4}{g_{\mu \nu }} - 4{{\cal U}_3}{{\cal K}_{\mu \nu }} + 12{{\cal U}_2}{\cal K}_{\mu \nu }^2 - 24{{\cal U}_1}{\cal K}_{\mu \nu }^3 + 24{\cal K}_{\mu \nu }^4) \nonumber\\
&&- \frac{{{c_5}}}{2}({{\cal U}_5}{g_{\mu \nu }} - 5{{\cal
U}_4}{{\cal K}_{\mu \nu }} + 20{{\cal U}_3}{\cal K}_{\mu \nu }^2 -
60{{\cal U}_2}{\cal K}_{\mu \nu }^3  + 120{{\cal U}_1}{\cal
K}_{\mu \nu }^4 - 120{\cal K}_{\mu \nu }^5) + ...
\end{eqnarray}
As a result, the gravitational and electromagnetic filed equations of the bulk theory are obtained as
\begin{equation} \label{gravitational equations}
{G_{\mu \nu }} + \Lambda {g_{\mu \nu }} + m_g^2{{\cal X}_{\mu \nu }} =  - \frac{1}{2}{g_{\mu \nu }}{\cal F} + 2{F_{\mu \lambda }}{F_\nu }^{\,\lambda },
\end{equation}
\begin{equation} \label{EM equations}
{\nabla _\mu }{F^{\mu \nu }} = 0.
\end{equation}
According to the variation of the bulk action (\ref{bulk
variation}) and asking for a well-defined variational principle,
one has to appropriately cancel the boundary terms by use of
adding surface term(s), ${\cal I}_{s}$. To do so, the
Gibbons-Hawking action, ${{\cal I}_{{\rm{GH}}}}$, can remove the
derivative terms of ${g_{\mu \nu }}$ normal to the boundary and is
given by
\begin{equation}
{{\cal I}_{{\rm{GH}}}} =  \frac{1}{{8\pi {G_d}}}\int_{\partial {\cal M}} {{d^{d - 1}}x\sqrt { - h} K},
\end{equation}
where $K$ is the trace of extrinsic curvature of boundary,
${\partial {\cal M}}$. On the other hand, the electromagnetic
boundary term has to be eliminated by use of imposing boundary
condition or proposing another surface term. There are two
possibilities which define the fixed potential and the fixed
charge ensembles at infinity as
\begin{eqnarray} \label{boundary conditions}
{\left. {\delta {A_\nu }} \right|_{\partial {\cal M}}} = 0&\longleftrightarrow & \rm{fixed\; potential\; ensemble}\nonumber\\
{\cal I}_s={\cal I}_{\rm{GH}}+{\cal I}_{EM}&\longleftrightarrow &\rm{fixed\; charge\; ensemble} \nonumber\\
\end{eqnarray}
where ${{\cal I}_{EM}}$ is a new surface term that is needed to
remove the electromagnetic boundary term in (\ref{bulk
variation}), and, consequently, to fix charge on the boundary,
with the following explicit form
\begin{equation}
{{\cal I}_{EM}} =  - \frac{1}{{4\pi {G_d}}}\int_{\partial {\cal M}} {{d^{d - 1}}x\sqrt { - h} {n_\mu }{F^{\mu \nu }}{A_\nu }}.
\end{equation}
We refer to the such gravitating systems with fixed charge and
fixed potential boundary conditions at infinity as the canonical
and the grand canonical ensembles, respectively, which is common
in literature. In our considerations, these boundary conditions
will be imposed separately in order to compare the results of
black holes' $PV$ criticality in both the canonical and the grand
canonical ensembles.

To find topological (AdS) black holes, we make use of the
following $d(=n+2)$-dimensional line element ansatz
\begin{equation} \label{metric ansatz}
d{s^2} =  - V(r)d{t^2} + \frac{{d{r^2}}}{{V(r)}} + {r^2}{h_{ij}}d{x_i}d{x_j}\,\,\,(i,j = 1,2,...,n),
\end{equation}
where the line element ${h_{ij}} d{x_i}d{x_j}$ is the metric of
$n$-dimensional (unit) hypersurface with the constant curvature
${d_{1}}{d_{2}k}$ and volume ${\omega _{n}}$ with the following
forms
\begin{equation}
{h_{ij}}d{x_i}d{x_j} = \left\{ \begin{array}{l}
dx_1^2 + \sum\limits_{i = 2}^{d - 2} {\prod\limits_{j = 1}^{i - 1} {{{\sin }^2}{x_j}dx_i^2 \qquad \qquad (k = +1)} } \\
\sum\limits_{i = 1}^{d - 2} {dx_i^2} \qquad \qquad \qquad \qquad \qquad \quad (k = 0)\\
dx_1^2 + {\sinh ^2}{x_1}\sum\limits_{i = 2}^{d - 2} {dx_i^2\prod\limits_{j = 2}^{i - 1} {{{\sin }^2}{x_j}} \,\,\,(k = -1)}
\end{array} \right.
\end{equation}
in which $\prod\limits_x^y {...} = 1$ if $x>y$. Another line
element ansatz is necessary for the reference metric ${f_{\mu \nu
}}$. We are primarily interested in building an effective field theory by use of the gravitational language which could describe some properties of different phases of matter. It has been indicated in a series of papers \cite{Vegh2013,Alberte2016SolidHolography,Alberte2018HolographicPhonons,Alberte2018BHelasticity} that the black hole solutions of Einstein gravity minimally coupled with a number of $N$ scalar fields (${\phi ^a}$) on AdS can describe different types of matter (such as solids and fluids) in a covariant way. In such theories the number of independent scalar fields (known as St\"uckelberg fields) is less than the number of spacetime dimensions $d$ (i.e., $N<d$). After the (unitary) gauge fixing  ${\phi ^a} = \delta _\mu ^a{x^\mu }$, the field structure of such theories are equivalent to a family of massive gravity on AdS with a singular and (spatial) degenerate reference metric, i.e., a reference metric with vanishing $tt$ and $rr$ entries on the diagonal. Remarkably as emphasized in Ref. \cite{Beekman2017PhysRep}, the holographic language of massive gravity with a singular reference metric in terms of St\"uckelberg fields  \cite{Alberte2013,Alberte2015,Alberte2016SolidHolography,Alberte2018HolographicPhonons,Alberte2018BHelasticity} is related to the language of gauge field theories for liquid crystals \cite{Beekman2017PhysRep,Beekman2017PRB}, which shows a deep connection between these theories and it becomes more manifest using BHC as speculated in Ref. \cite{DehghaniHendiMann2019}. On the other hand, in view of gauge/gravity duality, massive gravity on AdS with a singular (degenerate) reference metric is dual to homogenous and isotropic condensed matter systems which leads to a boundary theory with the finite DC conductivity \cite{Vegh2013,BlakeTong2013,Davison2013}, a desired property for normal conductors that is absent in massless gravities \cite{HHH2008PRL,HHH2008JHEP,Gregory2009GBsuperconductor,Gregory2010GBsuperconductor}. For these reasons, massive gravity with a singular reference metric is of direct interest for us. So we make use of the following singular and (spatial) degenerate ansatz \cite{Vegh2013, Cai2015massive}
\begin{equation} \label{reference metric}
{f_{\mu \nu }} = diag\left( {0,0,c_0^2{h_{ij}}} \right),
\end{equation}
in which $c_{0}$ is a positive constant. Since $f_{\mu \nu }$
depends only on the spatial components $h_{ij}$ of the spacetime
metric, the theory do not preserve general covariance in the
transverse spatial coordinates $x_{1}, x_{2},..., x_{d_2}$. This
choice of the reference metric establishes a subclass of dRGT
massive gravity theories, called as the reduced massive gravity
\cite{Alberte2013,Alberte2015}, that is ghost free and admits
exact black hole solutions on AdS \cite{Vegh2013, BlakeTong2013,
Davison2013, GhostFreeSingular2016}. 

Nevertheless, it is possible to find black hole solutions with a nonsingular reference metric (in which the $tt$ and $rr$ components of the reference metric are non-zero), but the analytic solutions can be found only for some specific values of massive coefficients (see Refs. \cite{Koyama2011PRL,Nieuwenhuizen2011PRD,GruzinovMirbabayi2011PRD,deRham2011BHs,ChargedBHs2013PRD,Babichev2014JHEP,Babichev2015CQG} for more details). The reference metric in these black hole solutions is assumed to be the Minkowski metric which means that diffeomorphism breaks along all of the temporal and spatial directions. Assuming the Minkowskian reference metric, spherically symmetric black hole solutions were found in Refs. \cite{Koyama2011PRL,Nieuwenhuizen2011PRD} and in the limit of vanishing graviton mass they go smoothly to the Schwarzschild and RN black holes on de Sitter space. Asymptotically flat black hole solutions were found in Ref. \cite{GruzinovMirbabayi2011PRD} and they are potentially considered as the viable classical solutions for stars and black holes in massive gravity, but the curvature diverges near the horizon of these solutions. In this regard, black hole solutions with non-singular horizon were introduced in Ref. \cite{deRham2011BHs}. There exist other interesting black hole solutions including charged and rotating ones that were the subject of Refs. \cite{ChargedBHs2013PRD,Babichev2014JHEP,Babichev2015CQG}. It should be noted that all of these solutions have been found in 4-dimensions. In addition, the author of \cite{TQDo2016a,TQDo2016b} found out Schwarzschild–Tangherlini–(A)dS in five-dimensional massive gravity as well as in massive bi-gravity with assumptions that the reference metric is compatible with (for massive gravity) or proportional (for massive bi-gravity) to the physical one.

Using the ansatz (\ref{reference metric}), the
interaction terms ${\cal U}_{i}$ are calculated as
\begin{equation} \label{potentials}
{{\cal U}_i} = {\left( {\frac{{{c_0}}}{r}} \right)^i}\prod\limits_{j = 2}^{i + 1} {{d_j}},
\end{equation}
in which we have used the convention $d_i=d-i$ (throughout this paper, this convention will be used). Considering the above relation, it is inferred that there are at
most $(d-2)$ potential terms (${\cal U}_{i}$) in a $d$-dimensional
spacetime and all the higher-order terms vanish identically.
Therefore, the upper bound ($d-2$) exists for summation in Eq.
(\ref{massive filed terms}) in all dimensions.

Solving the Maxwell field equations (\ref{EM equations}) yields
the $U(1)$ gauge field as
\begin{equation} \label{gauge potential}
{A_\mu} =\Big(\phi- \frac{q}{{{d_3}{r^{{d_3}}}}}\Big)\delta _\mu ^0,
\end{equation}
in which $q$ is a constant related to the total electric charge of
spacetime. The constant $\phi$ is obtained by the regularity
condition \cite{York1990,Nastase2015Book} at the horizon
($r=r_+$), i.e., $A_t(r_+) =0 $, which leads to $\phi =
\frac{q}{{{d_3}{r_ + ^{d_3}}}}$. The electric potential $\Phi$ of
the black hole spacetime is actually electrostatic potential
difference between the horizon and the boundary at infinity. We
choose the event horizon as the reference (i.e., $\Phi=0$ as
$r\rightarrow r_{+}$). Thus, the electric potential can be
measured at the infinity with respect to the horizon as
\begin{equation} \label{electric potential}
\Phi  = {\left. {{A_\mu }{\chi ^\mu }} \right|_{r \to \infty }} -
{\left. {{A_\mu }{\chi ^\mu }} \right|_{r \to {r_ + }}} =
\frac{q}{{{d_3}{r_ + ^{d_3}}}},
\end{equation}
where $\chi=\partial_{t}$ is the temporal Killing vector.
Moreover, the electric charge can be obtained using the Gauss' law
as
\begin{equation} \label{charge}
Q =\frac{{{\omega _{n}}}}{{4\pi }}q.
\end{equation}

Now, the gravitational field equations (\ref{gravitational
equations}) can be solved using the metric ansatz (\ref{metric
ansatz}). To do that, the $rr$-component of Eq.
(\ref{gravitational equations}) is enough for our purpose to
obtain the metric function $V(r)$. The $rr$-component of Eq.
(\ref{gravitational equations}) is given as
\begin{eqnarray}
{d_2}{d_3}\left( {k - f(r)} \right)& -& {d_2}r\left(
{\frac{{df(r)}}{{dr}}} \right) - 2\Lambda {r^2} - 2{q^2}{r^{ -
2{d_3}}} + m_g^2\sum\limits_{i = 1}^{d - 2}
{\Big({\frac{{c_0^i{c_i}}}{{r^{i - 2}}}\prod\limits_{j = 2}^{i +
1} {{d_j}} } \Big)} = 0.
\end{eqnarray}
Consequently, the metric function $V(r)$ is obtained as
\begin{eqnarray} \label{metric function}
V(r) &=& k - \frac{2 \Lambda {r^2}}{{d_1}{d_2}} -
\frac{m}{{{r^{{d_3}}}}} + m_g^2\sum\limits_{i = 1}^{d - 2} {\Big(
{\frac{{c_0^i{c_i}}}{{{d_2}{r^{i - 2}}}}\prod\limits_{j = 2}^i
{{d_j}} } \Big)} + \frac{{2{q^2}}}{{{d_2}{d_3}{r^{2{d_3}}}}},
\end{eqnarray}
where $m$ is a positive constant related to the finite mass of
spacetime. This solution is valid to all orders in arbitrary
dimensions, and, simultaneously, satisfies all components of Eq.
(\ref{gravitational equations}).

The existence of essential singularity of the spacetime is
confirmed by computing the Kretschmann scalar which is
\begin{eqnarray}
{R^{\alpha \beta \gamma \delta }}{R_{\alpha \beta \gamma \delta
}}& =& {\left( {\frac{{{\partial ^2}V(r)}}{{\partial {r^2}}}}
\right)^2} + 2{d_2}{\left( {\frac{1}{r}\frac{{\partial
V(r)}}{{\partial r}}} \right)^2}  + 2{d_2}{d_3}{\left(
{\frac{{V(r) - k}}{{{r^2}}}} \right)^2}.
\end{eqnarray}
Taking into account the metric function $V(r)$, it can be found
${R^{\alpha \beta \gamma \delta }}{R_{\alpha \beta \gamma \delta
}} \propto {r^{ - 4{d_2}}}$ near the origin (${r \to {0^ + }}$)
and, for $r \neq 0$, all the curvature scalars are finite. This
curvature singularity cannot be eliminated by any coordinate
transformation, while can be covered by an event horizon, $r_+$
($V(r_{+})=0$). The roots of the metric function $V(r)=g^{r r}=0$
specify the number of horizons, and, as reported in Refs.
\cite{Katsuragawa2015, HendiDehghani2019,Hendi2016JHEP}, the
metric function $V(r)$ could have more than two roots. So the
multi-horizon black hole solutions are found in massive gravity
and we assume that $r_+$ is the event horizon radius of the black
hole solutions, i.e., the largest real positive root of $V(r) =
0$.

The Hawking temperature of the black hole spacetimes can be
obtained by applying the definition of surface gravity
\cite{BardeenCarterHawking1973} or employing the Euclidean
formalism \cite{HawkingPage1983,Witten1998b}. Considering the
latter, by the analytic continuation of the Lorentzian metric
(\ref{metric ansatz}) to Euclidean signature, i.e., ${t_E} = it$,
we get
\begin{equation} \label{Euclidean metric}
ds_E^2 = V(r)dt_E^2 + \frac{{d{r^2}}}{{V(r)}} + {r^2}{h_{ij}}d{x_i}d{x_j}.
\end{equation}
The above Euclidean metric has a conical singularity at the
horizon ($r=r_{+}$), so regularity condition near the horizon
requires that the Euclidean time be periodic, i.e., ${t_E} \sim
{t_E} + \beta $ (otherwise the expansion of the Euclidean
spacetime around $r=r_{+}$ shows a conical singularity). Thus, one
obtains the Hawking temperature of the obtained TBHs in the
canonical ensemble as
\begin{eqnarray} \label{temperature-can}
\beta ^{ - 1}& =& T = {\left. {\frac{1}{{4\pi }}\frac{{\partial
V(r)}}{{\partial r}}} \right|_{r = {r_ + }}} ={\frac{{{d_2}{d_3}k
-2 \Lambda {r_ +^2} - 2{q^2}r_ + ^{ - 2{d_3}} +
m_g^2\sum\limits_{i = 1}^{d - 2} {\Big(
{\frac{{c_0^i\,{c_i}}}{{\,r_ + ^{i - 2}}}\prod\limits_{j = 2}^{i +
1} {{d_j}} } \Big)} }}{{4\pi {d_2}\,{r_ + }}}},
\end{eqnarray}
and, for the case of grand canonical ensemble (fixed $\Phi$), using $q=d_3 \Phi r_+^{d_3}$, it is given by
\begin{equation} \label{temperature-grand}
\beta ^{ - 1} = T = {\frac{{{d_2}{d_3}k -2 \Lambda {r_ +^2} - 2{d_3^2}{\Phi^2} + m_g^2\sum\limits_{i = 1}^{d - 2} {\Big( {\frac{{c_0^i\,{c_i}}}{{\,r_ + ^{i - 2}}}\prod\limits_{j = 2}^{i + 1} {{d_j}} } \Big)} }}{{4\pi {d_2}\,{r_ + }}}}.
\end{equation}

Working in the Euclidean formulation (for more details see Refs.
\cite{Klauber2013,Zee2010QFT,Natsuume2015AdS/CFT,Erdmenger2015}),
the semi-classical partition functions of TBHs may be evaluated
using the following path integral over the dynamical metric ($g
\equiv g_{\mu \nu}$) as
\begin{equation}
{\cal Z} = \int {{\cal D}[g,\varphi ]{e^{ - {{\cal I}_E}[g,\varphi ]}}},
\end{equation}
in which $\cal D$ denotes integration over all paths, $\varphi $
is considered as matter fields and ${\cal I}_E$ represents the
Euclidean version of the Lorentzian action ${\cal I}_G$ by
implementing the Wick rotation, ${t_E} = it$. In next sections, we
will explicitly evaluate semi-classical black hole partition
functions in both the canonical and the grand canonical ensembles
by employing the appropriate boundary conditions. This facilitates
the study of thermodynamics of TBHs.

\section{Black hole chemistry in canonical ensemble} \label{Canonical BHC}
\subsection{Canonical partition function} \label{partition function can}
The gravitational partition function in the canonical (fixed charge) ensemble is defined by following path integral
\begin{equation} \label{canonical parition function}
{{\cal Z}_{\rm{C}}} = \int {{\cal D}[g, A]} {e^{ - {{\cal I}_E}[g, A]}}\simeq e^{ - {{\cal I}_{on-shell}}(\beta,Q)},
\end{equation}
in which $A$ and ${\cal I}_{on-shell}$ represent the gauge field
($A_{\mu}$) and the on-shell gravitational action, respectively.
The most dominant contribution of the partition function
originates from substituting the classical solutions of the
action, i.e., the so-called on-shell action by applying the
saddle-point approximation. To do that, one has to first compute
the total on-shell action (${\cal I}_{on-shell}$) in the Euclidean
formalism. In the case of the canonical ensemble, we need to add
the electromagnetic surface term ${\cal I}_{EM}$ to the action,
i.e., ${{\cal I}_G} = {{\cal I}_b} +{{\cal I}_{GH}}+ {{\cal
I}_{EM}} + {{\cal I}_{ct}}$. Here, we compute the on-shell action
using the Hawking-Witten prescription (the so-called subtraction
method \cite{HawkingPage1983,Witten1998b}), and so, we only need
to evaluate the on-shell bulk action plus the electromagnetic
surface term (${\cal I}_{EM}$) as the associated boundary
condition with the fixed charge ensemble.

In order to have a finite gravitational partition function,
following Hawking-Witten prescription, we subtract the on-shell
action of the AdS background without black hole (referred to as
${\cal I}_{AdS}$), i.e., setting $m=Q=0$ in Eq. (\ref{metric
function}), from the on-shell action of the black hole spacetime
(referred to as ${\cal I}_{BH}$) and compute the explicit form of
on-shell action in Euclidean signature. That leads to free energy
difference as
\begin{equation}
F\equiv \Delta F = {\beta ^{ - 1}}\left( {{{\cal I}_{BH}} - {{\cal I}_{AdS}}} \right),
\end{equation}
in which the zero point energy of the boundary gauge theory (based
on AdS/CFT duality) is canceled. Now, we briefly explain how to
compute the on-shell action of charged TBHs in dRGT massive
gravity. First, the Ricci scalar ($R$) is obtained using the
gravitational field equations (\ref{gravitational equations}) as
\begin{equation}
R = \frac{1}{{{d_2}}}\left( {2 \Lambda d + 2 m_g^2 {\cal X}
+{d_4} {\cal F}} \right), \qquad {\cal X} \equiv {g^{\mu \nu
}}{{\cal X}_{\mu \nu }}.
\end{equation}
Using this equation and the explicit form of interaction
potentials (\ref{potentials}), the bulk Lagrangian presented in
(\ref{bulk action}) is explicitly calculated as
\begin{eqnarray}
{\cal L}_{bulk} &\equiv& R - 2\Lambda  - {\cal F} +
m_g^2\sum\limits_{i = 1}^{d - 2} {{c_i}{{\cal U}_i}(g,f)} =
\frac{2}{{{d_2}}}( {2 \Lambda  - {\cal F} )+ m_g^2\sum\limits_{i =
1}^{d - 2} {(i - 2)\frac{{c_0^i{c_i}}}{{{r^i}}}} \prod\limits_{j =
3}^{i + 1} {{d_j}} },
\end{eqnarray}
in which the following identity \cite{DehghaniHendiMann2019} has
been used
\begin{equation}
2\prod\limits_{j = 3}^{i + 1} {{d_j}}  + \sum\limits_{y = 1}^i
{{{( - 1)}^y} \frac{{i!}}{{(i - y)!}}} \prod\limits_{j = 2}^{i - y
    + 1} {{d_j}}  = (i - 2)\prod\limits_{j = 3}^{i + 1} {{d_j}},
\end{equation}
where $\prod\limits_x^y {...}  = 1$ if $x>y$. Substituting the
classical solutions, Eqs. (\ref{metric function}) and (\ref{gauge
potential}), the on-shell bulk action for the TBH spacetimes with
the fixed charge boundary condition can be computed after a long
and tedious calculation as
\begin{eqnarray} \label{BH action-can}
{{\cal I}_{BH}} &=& \frac{{\beta \omega _{n}}}{{16\pi
{G_d}}}\Bigg[\frac{2}{{{\ell ^2}}}{r^{d_1}} - m_g^2\sum\limits_{i
= 1}^{d - 2} {\frac{{(i - 2)c_0^i{c_i}}}{{d - i - 1}}{r^{d - i -
1}}\prod\limits_{j = 3}^{i + 1} {{d_j}} } -
\frac{{4{q^2}}}{{{d_2}r ^{{d_3}}}} \Bigg]_{{r_ + }}^{R},
\end{eqnarray}
in which "$R$" is an upper cutoff on the radial integrations in
order to regularize the action, and will be canceled at the end.
In the above calculation, we have used the electromagnetic surface
term, ${\cal I}_{EM}$, in the Euclidean signature as
\begin{eqnarray} \label{EM surface term}
{{\cal I}_{EM}}& =&  - \frac{1}{{4\pi {G_d}}}\int_{\partial {\cal
M}} {{d^{d - 1}}x\sqrt { h} {n_\mu }{F^{\mu \nu }}{A_\nu }}=
\frac{{\beta {\omega _{n}}}}{{16\pi {G_d}}}\Bigg[
{\frac{{4{q^2}}}{{{d_3}r_ + ^{{d_3}}}}} \Bigg].
\end{eqnarray}
Repeating the same procedure for the AdS background in massive gravity (without any matter or electromagnetic field, i.e., setting $Q=m=0$), one obtains
\begin{eqnarray} \label{AdS action}
{{\cal I}_{AdS}} = \frac{{{\beta}_{0} \omega _{n}}}{{16\pi {G_d}}}\Bigg[ {\frac{2}{{{\ell ^2}}}{R^{d_1}} - m_g^2\sum\limits_{i = 1}^{d - 2} {\frac{{(i - 2)c_0^i{c_i}}}{{d - i - 1}}{R^{d - i - 1}}\prod\limits_{j = 3}^{i + 1} {{d_j}} } } \Bigg], \nonumber\\
\end{eqnarray}
with the period ${\beta}_0$. Both the AdS and the black hole
spacetimes at $r = R$ must have the same geometry, i.e. ${\beta
_0}V_0(R)^{1/2} = \beta V{(R)^{1/2}}$, which leads to
\begin{equation}
{\beta _0} = \beta \left( {1 - \frac{{m{\ell ^2}}}{{2{r^{d - 1}}}} + O({r^{ - 2(d - 1)}})} \right).
\end{equation}
Using this fact and the following identity
\cite{DehghaniHendiMann2019}
\begin{equation}
\frac{1}{{{d_2}}}\prod\limits_{j = 2}^i {{d_j}}  + \frac{{i - 2}}{{d - i - 1}}\prod\limits_{j = 3}^{i + 1} {{d_j}}  = (i - 1)\prod\limits_{j = 3}^i {{d_j}},
\end{equation}
the renormalized on-shell action in the canonical ensemble is
eventually computed as
\begin{eqnarray} \label{canonical action}
{{\cal I}_{on - shell}} &\equiv& \mathop {\lim }\limits_{R \to
\infty } \left( {{{\cal I}_{{\rm{BH}}}} - {{\cal I}_{{\rm{AdS}}}}} \right) \nonumber \\
&=& \frac{{\beta \omega _{n} r_ + ^{{d_3}}}}{{16\pi
{G_d}}}\Bigg[k - \frac{{r_ + ^2}}{{{\ell ^2}}} + \frac{{2(2d -
5)}q^2}{{{d_2}{d_3}r_ + ^{2{d_3}}}}+ m_g^2\sum\limits_{i = 1}^{d -
2} {\Big( {\frac{{(i - 1)c_0^i{c_i}}}{{r_ + ^{i -
2}}}\prod\limits_{j = 3}^i {{d_j}} } \Big)} \Bigg].
\end{eqnarray}

\subsection{Thermodynamics} \label{Can Thermodynamics}

Thermodynamic quantities are straightforwardly extracted form the
obtained partition function. Using the canonical partition
function, Eqs. (\ref{canonical parition function}) and
(\ref{canonical action}), the mass of the black hole is computed
as (setting $G_d=1$)
\begin{eqnarray} \label{Mass-can}
M &=&  - \frac{\partial }{{\partial \beta }}\ln {{\cal
Z}_{\rm{C}}} = \frac{{{d_2}{\omega _{n}}}}{{16\pi }}r_ +
^{{d_3}}\Bigg[ k + {{\left( {\frac{{{r_ + }}}{\ell }} \right)}^2}
+ m_g^2\sum\limits_{i = 1}^{d - 2} {\Big(
{\frac{{c_0^i{c_i}}}{{{d_2}r_ + ^{i - 2}}}\prod\limits_{j = 2}^i
{{d_j}} } \Big)} + \frac{{2{q^2}}}{{{d_2}{d_3}r_ + ^{2{d_3}}}}
\Bigg],
\end{eqnarray}
where, in the extended phase space ($\Lambda=-\frac{{d_1}{d_2}}{2
\ell ^2}=-8 \pi P$), has to be interpreted as the black hole
enthalpy, $M\equiv H$. The above black hole mass is in agreement
with the ADM mass formula as
\begin{equation}
M = \frac{{d_2}{\omega_{n}}}{16\pi }m,
\end{equation}
in which the constant $m$ is obtained from $V({r_ + }) = 0$ (see
Eq. \ref{metric function}), and thus, the same result as Eq.
(\ref{Mass-can}) is obtained. Working in the extended phase space,
the free energy in the canonical ensemble is obtained as
\begin{eqnarray} \label{Gibbs free energy}
G& =& {\beta ^{ - 1}}\ln {{\cal Z}_{\rm{C}}}(T,P,Q) = M - TS \nonumber \\
&=&\frac{{{\omega _{{n}}}r_ + ^{{d_3}}}}{{16\pi }}\Bigg[k -
\frac{{16\pi Pr_ + ^2}}{{{d_1}{d_2}}} + \frac{{(4d -
10){q^2}}}{{{d_2}{d_3}r_ + ^{2{d_3}}}} + m_g^2\sum\limits_{i =
1}^{d - 2} {\Big( {\frac{{(i - 1)c_0^i{c_i}}}{{r_ + ^{i -
2}}}\prod\limits_{j = 3}^i {{d_j}} } \Big)} \Bigg],
\end{eqnarray}
which, in fact, is the Gibbs free energy of the AdS black hole.
Using the obtained free energy, the other thermodynamic variables
(i.e. $V$, $\Phi$ and $S$) can be easily computed (notice that
$r_+$ is understood as a function of $P$ and $T$ according to Eq.
(\ref{temperature-can})). The thermodynamic volume is given by
\begin{equation} \label{volume}
V = {\left( {\frac{{\partial G}}{{\partial P}}} \right)_{T,Q}} = \frac{{{\omega _{n}}}}{{{d_1}}}r_ + ^{{d_1}}.
\end{equation}
The electric potential can be calculated using the following thermodynamic relation as
\begin{equation}
\Phi  = {\left( {\frac{{\partial G}}{{\partial Q}}} \right)_{T,P}} = \frac{q}{{{d_3}r_ + ^{{d_3}}}},
\end{equation}
which is in agreement with the Eq. (\ref{electric potential}). Finally, the black hole entropy is given by
\begin{equation} \label{entropy}
S =  - {\left( {\frac{{\partial G}}{{\partial T}}} \right)_{P,Q}} = \frac{{{\omega _n}}}{4}r_ + ^{{d_2}},
\end{equation}
which satisfy the so-called area law and is in agreement with the relation $S=\beta (M-G)$ as expected.

In conclusion, the obtained thermodynamic quantities satisfy
analytically the first law of thermodynamics in the Gibbs energy
representation, i.e. $dG=-SdT+\Phi dQ+VdP$. Furthermore, one can
use the Legendre transform $G=M-TS$ to write down the first law in
the enthalpy representation as $dM=TdS+\Phi dQ+VdP$, in which the
temperature and thermodynamic volume are respectively obtained
using $T = {\left( {{{\partial M}}/{{\partial S}}} \right)_{P,Q}}$
and $V = {\left( {{{\partial M}}/{{\partial P}}} \right)_{S,Q}}$
in agreement with Eqs. (\ref{temperature-can}) and (\ref{volume}).
Finally, using these ingredients, it can be derived that obtained
thermodynamic quantities obey the extended Smarr formula as
\begin{equation} \label{Smarr}
(d - 3)M = (d - 2)TS + (d - 3)\Phi Q - 2PV + \sum\limits_{i = 1}^{d - 2} {(i - 2){{\cal C}_i}{c_i}},
\end{equation}
where the conjugate potentials (${{\cal C}_i}$) corresponding to
the massive couplings ($c_i$) are given by
\begin{equation}
{{\cal C}_i} = {\left( {\frac{{\partial M}}{{\partial {c_i}}}} \right)_{S,P,Q,{c_{j \ne i}}}} = \frac{{{\omega _{{n}}}}}{{16\pi }}m_g^2c_0^ir_ + ^{{d_{i + 1}}}\prod\limits_{j = 2}^i {{d_j}}.
\end{equation}
As a result, the extended Smarr relation suggests that one should
take into account $c_i$'s as the new thermodynamic variables. This
leads to the extended first law as $dM=TdS+\Phi
dQ+VdP+\sum\limits_{i = 1}^{d - 2} {{{\cal C}_i}d{c_i}}$.
Comparing Eq. (\ref{Smarr}) with Refs.
\cite{Kastor2009CQG,Kastor2010LovelockSmarr}, the same result for
the Smarr formula in the extended phase space is obtained if one
invokes the method of scaling argument as
\begin{eqnarray}
(d - 3)M &=& (d - 2)\left( {\frac{{\partial M}}{{\partial S}}}
\right)S + (d - 3)\left( {\frac{{\partial M}}{{\partial Q}}}
\right)Q - 2\left( {\frac{{\partial M}}{{\partial P}}} \right)P +
\sum\limits_{i = 1}^{d - 2} {(i - 2)\left( {\frac{{\partial M}}
{{\partial {c_i}}}} \right){c_i}},
\end{eqnarray}
with the following scalings for the thermodynamic quantities
\begin{eqnarray}
&& \left[ M \right] = {L^{d - 3}},\,\,\left[ {{c_i}} \right] =
{L^{i - 2}},\,\,\left[ P \right] = {L^{ - 2}}, \,\,\,\, \left[ S
\right] = {L^{d - 2}},\,\,\left[ Q \right] = {L^{d - 3}}.
\end{eqnarray}

\subsection{Holographic phase transitions}
Using Eq. (\ref{temperature-can}), the canonical equation of state of charged TBHs is calculated in the extended phase space as
\begin{eqnarray} \label{Can EOS}
P& = &\frac{{{d_2}\tilde T}}{{4{r_ + }}} - \frac{{{d_2}{d_3}}
k_{\rm {eff}}^{(\rm {C})}}{{16\pi r_ + ^2}} -
\frac{{m_g^2}}{{16\pi }}\sum\limits_{i = 3}^{d - 2} {\Big(
{\frac{{c_0^i{c_i}}}{{\,r_ + ^i}}\prod\limits_{j = 2}^{i + 1}
{{d_j}} } \Big)} + \frac{{{q^2}}}{{8\pi r_ + ^{2{d_2}}}},
\end{eqnarray}
in which the effective topological factor $k_{\rm {eff}}^{(\rm
{C})}$ has been introduced as
\begin{equation}
k_{\rm {eff}}^{(\rm {C})} \equiv [k + m_g^2c_0^2{c_2}],
\end{equation}
and $\tilde T$ is the shifted Hawking temperature
\cite{PVmassive2015PRD,Reentrant-dRGTmassive-2017,HendiDehghani2019,DehghaniHendiMann2019}
with the following explicit form
\begin{eqnarray} \label{shifted Hawking temp}
\tilde T &=& T - \frac{{m_g^2{c_0}{c_1}}}{{4\pi }} =
\frac{{{d_2}{d_3}k - 2\Lambda r_ + ^2 - 2{q^2}r_ + ^{ - 2{d_3}} +
m_g^2\sum\limits_{i = 2}^{d - 2} {\Big( {\frac{{c_0^i{c_i}}}{{r_ +
^{i - 2}}}\prod\limits_{j = 2}^{i + 1} {{d_j}} } \Big)} }}{{4\pi
{d_2}{r_ + }}}.
\end{eqnarray}

The inflection point(s) of isothermal curves in $P-v$ diagrams determine the critical point(s), i.e., by applying the following relations
\begin{eqnarray} \label{inflection points}
{\left( {\frac{{\partial P}}{{\partial v}}} \right)_T} = 0\,\,\, &\Longleftrightarrow& \,\,\,{\left( {\frac{{\partial P}}{{\partial {r_ + }}}} \right)_T} = 0, \nonumber\\
{\left( {\frac{{{\partial ^2}P}}{{\partial {v^2}}}} \right)_T} = 0\,\,\, &\Longleftrightarrow& \,\,\,\left(\frac{{{\partial ^2}P}}{{\partial {r_ +^2}}} \right)_T= 0,
\end{eqnarray}
where the specific volume $v$ is proportional to $ r_ + $ as $v  = 4{r_ + }\ell
_{\rm{P}}^{{d_2}}/{d_2}$ \cite{KubiznakMann2012,Gunasekaran2012Mann,PVmassive2015PRD,Reentrant-dRGTmassive-2017,HendiDehghani2019,DehghaniHendiMann2019}. So, because of the dependency between $r_+$, $v$ and the thermodynamic volume $V$, criticality in one of $P-r_{+}$, $P-v$ or $P-V$ planes indicates criticality in the others. On the other hand, since the essential information of the phase transitions encodes in the $G-T$ diagrams, so we mainly discuss based on the isobaric curves of these diagrams in the next sections.

Implementing Eq. (\ref{inflection points}) for the equation of state (\ref{Can EOS}) leads to
\begin{eqnarray} \label{critical points Can}
{d_3} k_{\rm{eff}}^{\rm{(C)}} r_ + ^{2{d_3}} &+&
m_g^2\sum\limits_{i = 3}^{d - 2} {\Big( {i(i - 1)c_0^i{c_i}r_ +
^{2d - i - 4}\prod\limits_{j = 3}^{i + 1} {{d_j}} } \Big)} - 2(2d
- 5){q^2} = 0.
\end{eqnarray}
As will be shown, the number of physical critical point(s) of the
above equation specifies the type of phase transition. In the
following, we will analyze the canonical ensemble holographic
phase transitions in massive gravity theory case by case with
detail. In advance, it should be noted that our considerations are
for all types of charged TBHs (with $k=0,\pm1$). In fact,
according to the given equations in this section, the combination
$k_{\rm{eff}}^{\rm{(C)}} \equiv {[k + {m_g^2}c_0^2{c_2}]}$ as an
effective topological factor always appears. If we find a set of
parameters related to a critical behavior in a charged-AdS black
hole system with a specific event horizon geometry, one can always
obtain the same critical behavior in another black hole system
with different horizon geometry. The only necessity is that the
same value must be provided for $k_{\rm{eff}}^{\rm{(GC)}}$, which
is always possible by varying the massive constant $c_2$.
Consequently, the same critical points with the same critical
behavior are found for the case of the spherical, planar, and
hyperbolic black holes. This is a remarkable property of TBHs in
massive gravity (first indicated in \cite{DehghaniHendiMann2019}).

\subsubsection{van der Waals (vdW) phase transition}
The vdW phase transition in the context of massive gravity is
widely discussed before in Refs.
\cite{Cai2015massive,PVmassive2015PRD,Hendi2017Mann-PRD}. But
here, we try to generalize the analytical results to higher
dimensions. To have the vdW phase transition, one (physical)
critical point is needed. This phenomenon can commence to appear
in $d \ge 4$ dimensions. In a 4-dimensional spacetime, only the
first two massive couplings ($c_1$ and $c_2$) are present. So, we
can simply assume that the first two massive couplings are
non-zero and the others vanish in higher dimensions. Of course,
according to Eq. (\ref{critical points Can}), one can always find
one (physical) critical point associated with the vdW behavior by
use of an appropriate fine tuning of massive parameters. Here,
using Eq. (\ref{critical points Can}), we apply the first approach
which leads to the following critical point equation
\begin{equation}
{d_3} k_{\rm{eff}}^{\rm{(C)}} r_ + ^{2{d_3}} - 2(2d - 5){q^2} = 0.
\end{equation}
The critical horizon radius is obtained easily as
\begin{equation} \label{vdW critical radius-can}
{r_c} = {\left( {\frac{{2(2d - 5){q^2}}}{{{d_3} k_{\rm{eff}}^{\rm{(C)}}}}} \right)^{\frac{1}{{2{d_3}}}}},
\end{equation}
in which the constraint $k_{\rm{eff}}^{\rm{(C)}}>0$ must be
satisfied. According to the later constraint, there is no
limitation on the value of the $U(1)$ charge in Einstein ($m_g=0$)
or massive gravity. Evidently, in the massless limit of gravitons
($m_g=0$) which leads to the Einstein gravity, no phase transition
and critical behavior take place for charged TBHs with Ricci flat
or hyperbolic horizon geometries (i.e., $k=0,-1$).

Regarding Eq. (\ref{vdW critical radius-can}), the critical pressure and temperature are computed as
\begin{equation}
{P_C} = \frac{{d_3^2 k_{\rm{eff}}^{\rm{(C)}}}}{{16\pi r_c^2}},
\end{equation}
and
\begin{equation} \label{critical temp-can}
{\tilde T_C} = \frac{{d_3^2 k_{\rm{eff}}^{\rm{(C)}}}}{{(2d - 5)\pi {r_c}}}.
\end{equation}

\begin{figure}[tbp]
    \epsfxsize=7cm \epsffile{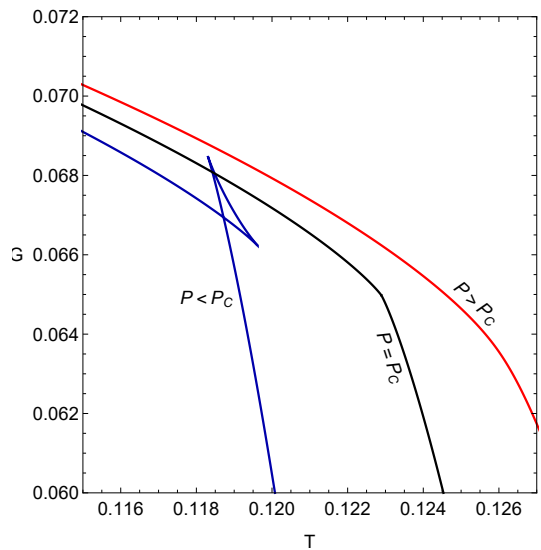}

    \caption{\textbf{vdW phase transition in the canonical (fixed charge) ensemble:} $G-T$ diagram for a spherical black hole with [$k=+1$, $c_2=0$], or a planar black hole with [$k=0$, $c_2=1$], or a hyperbolic black hole with [$k=-1$, $c_2=2$]. The other parameters have been set as $d=4$, $m_g=1$, $c_{0}=1$, $c_{1}=1$, and $q=1$ \\
        \textit{Critical data}: ($T_C=0.122894$, $P_C=0.003316$)}

    \label{GT_vdW_can}
\end{figure}

Now, using these critical quantities, we can easily plot the vdW
behavior of phase transition for a set of TBHs. In Fig.
(\ref{GT_vdW_can}), $G-T$ diagrams for the isobaric curves near
the critical point are depicted. As seen, for the range $P<P_C$,
the characteristic swallowtail form is observed which closely
resembles the vdW phase transition in fluids. For the range
$P>P_C$, the isobars correspond to the ideal gas with a single
phase. Moreover, the presented example is a generic feature of all
types of TBHs.

Using ${v} = {4r_+}/{d_2}$ in the geometric units, the obtained
thermodynamic quantities at the critical point satisfy the
following universal ratio
\begin{equation} \label{ratio-can-shifted}
\frac{{{P_C}{r_c}}}{{{{\tilde T}_C}}} = \frac{{2d - 5}}{{16}}\Longleftrightarrow \frac{{{P_C}{v_c}}}{{{{\tilde T}_C}}} = \frac{{2d - 5}}{{4{d_2}}},
\end{equation}
in which the shifted Hawking temperature ($\tilde T = T
-{{m_g^2{c_0}{c_1}}}/{{4\pi }}$) is used and the constraint
$k_{\rm{eff}}^{\rm{(C)}}>0$ which imposed from Eq. (\ref{vdW
critical radius-can}) can ensure the positivity of ${\tilde T}_C$.
Interestingly, the obtained critical ratio does not depend on the
topology of the event horizon. This exactly matches with the
universal ratio of RN-AdS black holes in Einstein gravity, and in
$d=4$ dimensions, the universal ratio of vdW fluid, i.e., $3/8$,
is obtained too. Obviously, in terms of the standard Hawking
temperature ($T$), the universal ratio will be a function of
graviton's mass ($m_g$). Regarding this case, one finds the
standard universal ratio as
\begin{equation} \label{ratio-can}
\frac{{{P_C}{v_c}}}{{{T_C}}} = \frac{{(2d - 5)d_3^2 k_{\rm{eff}}^{\rm{(C)}}}}{{{d_2}\left( {4d_3^2 k_{\rm{eff}}^{\rm{(C)}}+ (2d - 5)m_g^2{c_0}{c_1}{r_c}} \right)}}.
\end{equation}
It is interesting to note that expanding the above universal ratio
around the infinitesimal values of the graviton's mass ($m_g$)
yields
\begin{eqnarray}
\frac{{{P_C}{v_c}}}{{{T_C}}} = \frac{{2d - 5}}{{4{d_2}}} &\Big \{
&1 - \frac{{(2d - 5)m_g^2{c_0}{c_1}}}{{4d_3^2k}}{{\left(
{\frac{{2(2d - 5){q^2}}}{{{d_3}k}}} \right)}^{\frac{1}{{2{d_3}}}}}
+ O(m_g^4) \Big \},
\end{eqnarray}
which only is true for the case of spherical symmetry ($k=+1$).
According to Eq. (\ref{vdW critical radius-can}), when the
geometry of event horizon is planar or hyperbolic (i.e., $k=0,
-1$), the zero limit ($m_g \to 0$) of critical radius does not
exist anymore and thus one is not allowed to expand the universal
ratio ${P_C}{v_c}/{T_C}$ around $m_g=0$. This shows the drastic
effect of event horizon's geometry in the massless limit ($m_g=0$)
of massive gravity as expected. It is natural since we know that
there is no criticality for planar or hyperbolic black holes in
Einstein gravity. Consequently, since the expansion around $m_g=0$
is not permissible (for $k=0,-1$), it can be inferred from Eq.
(\ref{vdW critical radius-can}) that a lower mass bound is needed
to have a positive definite critical radius and the subsequent
critical behavior for Ricci flat or hyperbolic black holes.

\subsubsection{Reentrant phase transition (RPT)} \label{RPT-can}

We now discuss the reentrant behavior of phase transition (first
seen in \cite{Gunasekaran2012Mann,Altamirano2012Reentrant}) in the
canonical ensemble. As stated in \cite{HendiDehghani2019}, the RPT
phenomenon can appear when the equation of state could give birth
to two critical points in which the associated pressures and
temperatures are positive definite, but, it is seen that only one
of these critical points, referred to as ($T_C$, $P_C$), is
physical and the other one is unphysical since it cannot minimize
the Gibbs free energy. By further studying the phase space, it
will be evident that a virtual triple point ($T_{Tr}$, $P_{Tr}$)
and another critical point ($T_{Z}$, $P_{Z}$) emerge. As a result,
three separate phases of black holes appear. The critical points ($T_{Z}$,
$P_{Z}$) and ($T_C$, $P_C$) are, respectively, the endpoints of
the zeroth-order and first order coexistence lines. Moreover, the
first order and the zeroth-order coexistence lines join at the
virtual triple point ($T_{Tr}$, $P_{Tr}$).

In the context of massive gravity, it seems that the phenomenon of
RPT may be observed when higher order interactions of massive
gravitons up to the third interaction terms are considered (which
is possible in $d \ge 5$ dimensions). But we speculate this is not
the case. Using Eq. (\ref{critical points Can}), the equation of
critical point(s) in five-dimensional spacetime reads (setting
$c_0=1$)
\begin{equation} \label{5D critical point eq Can}
 m_g^2 c_2 {r_+^4}+3 m_g^2 c_3 {r_+^3} - 5 q^2=0,
\end{equation}
in which, without loss of generality, it is assumed that $k=0$ (so
we have $k_{\rm{eff}}^{\rm{(C)}}=m_g^2 c_0^2 c_2$). Also, $m_g^2$
can be absorbed into the massive couplings $c_1$ and $c_2$.
Analyzing Eq. (\ref{5D critical point eq Can}) by Mathematica
Software \cite{Mathematica} shows that this equation could possess
two critical points according to the following conditions
\begin{equation} \label{5D condition}
0<q^2<- \frac{3^7}{5 \times 2^8} \frac{c_3^4}{c_2^3},\quad {c_2}<0,\quad {c_3}>0,
\end{equation}
in which it is assumed that $r_+>0$. But, the pressures
corresponding to the obtained critical roots must be positive
definite, i.e., $P>0$, and this later condition cannot be
satisfied for both roots simultaneously. In fact, by combining Eq.
(\ref{5D critical point eq Can}) with the condition $P>0$, a
critical radius is obtained with the following fine tuning
\begin{equation}
0<q^2<-\frac{3^4}{5 \times 2^4} \frac{c_3^4}{c_2^3}, \quad{c_2}<0, \quad {c_3}>0,
\end{equation}
and another one can also be found as
\begin{equation}
c_2>0, c_3>0, \quad \text{or} \quad c_2>0, c_3<0.
\end{equation}
Examining the above conditions, it is seen that the first root
requires $c_2<0$ while the second root requires $c_2>0$. So there
are no combinations of the parameters at the same time where both
roots are solutions. Thus, the phenomenon of RPT cannot take place
in $d=5$ dimensions. Now, following this procedure, we can discuss
RPT in 6-dimensions. As summarized in table \ref{tab:sign
variations}, in a six-dimensional spacetime, there are eight
possibilities for the sign variations of massive coupling
constants $c_2$, $c_3$ and $c_4$. Only in the cases of (1) and
(5), two roots can be found. But, the same as RPT in
five-dimensions, one cannot find any region in the phase space
where both roots and the associated pressures are positive
definite (but, interestingly, a region with three physical
critical points are found which is the subject of Sect.
\ref{vdW-type PT}). Thus, in 6-dimensions, the RPT phenomenon does
not take place in the canonical ensemble of massive gravity's
charged TBHs. This analytical procedure does not work in the case
of higher dimensions, i.e., $d \ge 7$, since the critical point
equation has mathematically a more complicated structure. So we
applied numerical analysis and did not find any evidence that
shows spacetime dimensions with the range $d=7,8,9$ can exhibit
reentrant behavior for phase transition as well. Motivated by
this, we speculate that charged TBHs may not exhibit RPT in the
canonical ensemble of charged TBHs in massive gravity (at least in
$d=4,5,6$ and possibly in $d=8,9$ dimensions). Whether or not such
phenomenon exists for massive gravity's TBHs in higher dimensions
($d \ge 7$) remains an open question.

\begin{table}
    \centering
    \caption{The sign variations in $d=6$ and the associated critical roots (points) of Eq. (\ref{critical points Can}). In this table, by "physical root" we mean that both the temperature and pressure associated with the critical root are positive definite.}
    \label{tab:sign variations}
    \begin{tabular*}{\columnwidth}{@{\extracolsep{\fill}}llllll@{}}
        \hline
        {case} & $c_2$ & $c_3$ & $c_4$ & roots & physical roots
        \\
        \hline
        \, (1) & $+$ & $+$ & $+$ &\, 2& \qquad 1  \\
        \, (2) & $+$ & $+$ & $-$ &\, 1 & \qquad 1  \\
        \, (3) & $+$ & $-$ & $+$ &\, 3 & \quad 1 or 3  \\
        \, (4) & $+$ & $-$ & $-$ &\, 1 &  \qquad 1  \\
        \, (5) & $-$ & $+$ & $+$ &\, 2 & \qquad 1  \\
        \, (6) & $-$ & $+$ & $-$ &\, 1 & \qquad 1  \\
        \, (7) & $-$ & $-$ & $+$ &\, 1 & \qquad 1  \\
        \, (8) & $-$ & $-$ & $-$ &\, 0 & \qquad 0  \\
        \hline
    \end{tabular*}
\end{table}

\subsubsection{Triple point and small/intermediate/large black hole (SBH/IBH/LBH) phase transition}

\begin{figure}[tbp]
    \epsfxsize=8cm \epsffile{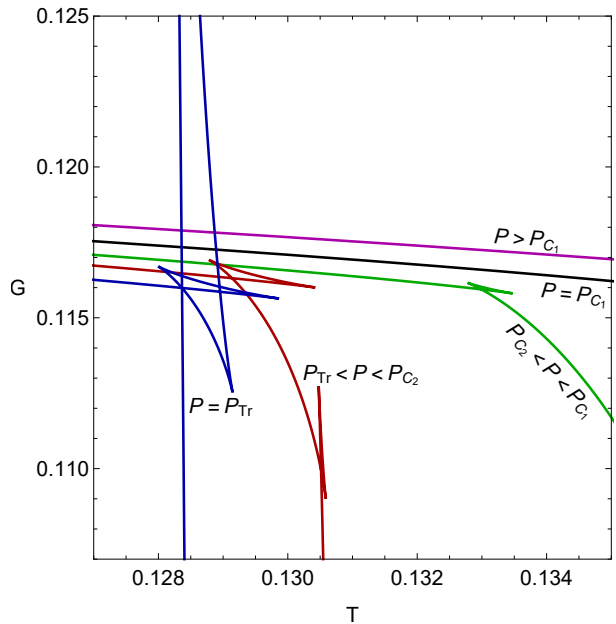}

    \caption{\textbf{Triple point in the canonical (fixed charge) ensemble:} $G-T$ diagram for a spherical black hole with [$k=+1$, $c_2=-0.1$], or a planar black hole with [$k=0$, $c_2=0.9$], or a hyperbolic black hole with [$k=-1$, $c_2=1.9$]. The other parameters have been set as $d=6$, $m_g=1$, $c_{0}=1$, $c_{1}=1$, $c_{3}=-0.8$, $c_{4}=0.5$ and $q=0.8$ \\
        \textit{Critical data}: ($T_{C_{1}}=0.138126$, $P_{C_{1}}=0.011428$), ($T_{C_{2}}=0.131100$, $P_{C_{2}}=0.004501$) and ($T_{Tr}=0.128365$, $P_{Tr}=0.003801$)}

    \label{GT_Triple_can}
\end{figure}

Here, we present the first explicit demonstration of the triple
point, typical of many materials in nature, in charged TBHs with
massive gravitons which takes place in $d \ge 6$ dimensions. The
analogue of triple point in the neutral black holes of massive
gravity was reported in \cite{DehghaniHendiMann2019}, in which
this phenomenon takes place in dimensions with the range $d \ge
7$, and necessarily one needs to consider up to the five graviton
self-interaction potentials. But here, in the canonical ensemble,
we explicitly show that this critical behavior can be observed for
charged TBHs in spacetimes with $d \ge 6$ in which only the first
four graviton self-interaction terms are present. According to the
table \ref{tab:sign variations}, in order to have the analogue of
triple point in massive charged TBHs in the canonical ensemble, we
have to consider the sign variations presented in the case (3). We
assume that the first four potential terms are nonzero and the
rest of them vanish, i.e., $c_i=0$ for $i \ge 5$. This leads to
the following critical point equation
\begin{eqnarray} \label{CPE-triple-can}
&&{d_3} k_{\rm{eff}}^{\rm{(C)}} r_ + ^2 +
3{d_3}{d_4}m_g^2c_0^3{c_3}{r_ + } +
6{d_3}{d_4}{d_5}m_g^2c_0^4{c_4}- 2(2d - 5){q^2}r_ + ^{ - 2{d_4}} =
0,
\end{eqnarray}
which predicts three physical critical points. By physical we just
mean that the associated pressures and temperatures are positive
definite (but always, one of those critical points cannot minimize
the Gibbs free energy). This is shown in Fig. \ref{GT_Triple_can},
where we have depicted the $G-T$ diagram for various isobaric
curves. For pressures in the range $P<P_{C_1}$, we first observe
the swallowtail (vdW) behavior which indicates the first-order
phase transition. This is the transition from the LBH region to
the SBH one. Then, for the isobars with $P_{Tr} < P < P_{C_2}$,
two swallowtails indicating the appearance of two first-order
phase transitions is observed which implies three-phase behavior.
This is the SBH/IBH/LBH phase transition that resembles the
solid/liquid/gas phase transition in usual substances. Finally, as
displayed in Fig. \ref{GT_Triple_can}, the two swallowtails
eventually merge at the gravitational triple point ($T_{Tr}$,
$P_{Tr}$) by further decreasing the pressure.

\subsubsection{vdW type phase transition} \label{vdW-type PT}
As stated in the previous section, the equation of state
(\ref{CPE-triple-can}) could possess three critical points in
which the associated pressures and temperatures are positive
definite. The three-phase behavior associated with the triple
point takes place when two critical points, referred to as
($T_{C_1}$, $P_{C_1}$) and ($T_{C_2}$, $P_{C_2}$), can minimize
the Gibbs free energy but the third one cannot. Interestingly,
another critical phenomenon can happen when two critical points
cannot minimize the Gibbs free energy but the other critical point
can do this. In fact, this situation had already appeared in Sect.
\ref{RPT-can}, the case (3) for the sign variations of massive
coefficients in table \ref{tab:sign variations}. This kind of
phase structure yields a critical behavior which here is referred
to as vdW type phase transition and can be obtained by use of
varying the electric charge of TBHs. This situation is similar to
that seen in Ref. \cite{Wei2014} for the spherically symmetric
charged-AdS black holes (in the canonical ensemble) within the
framework of Gauss-Bonnet gravity, exactly in $d=6$ dimensions. We
confirm that this phenomenon happens in the context of massive
gravity as well, and, interestingly it starts to appear in $d \ge
6$ dimensions for all types of TBHs.\footnote{It should be noted
that the authors of Ref. \cite{Wei2014} have studied the $P-v$
criticality of charged black holes for the case of $\alpha_{GB}=1$
, which is very limited. As they stated, in the case of arbitrary
Gauss-Bonnet coupling constant, one may find this phenomenon in
higher dimensions as well.}

\begin{figure}[tbp]
    \epsfxsize=8cm \epsffile{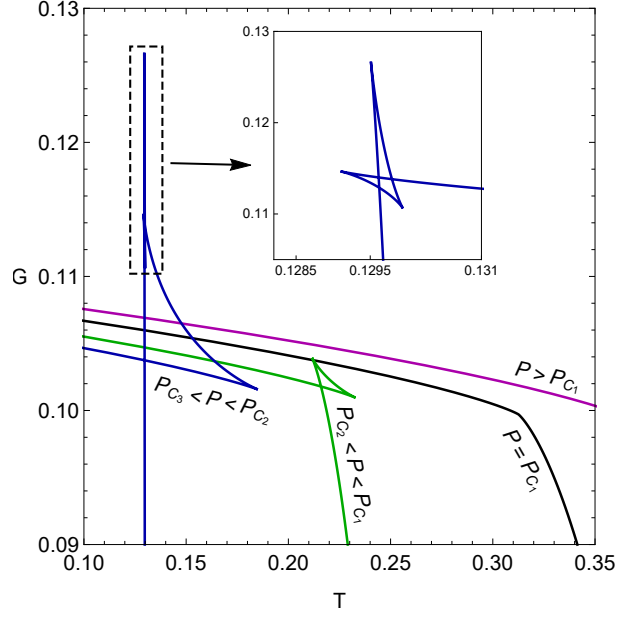}

    \caption{\textbf{vdW type phase transition in the canonical (fixed charge) ensemble:} the $G-T$ diagram for a spherical black hole with [$k=+1$, $c_2=-0.1$], or a planar black hole with [$k=0$, $c_2=0.9$], or a hyperbolic black hole with [$k=-1$, $c_2=1.9$]. The other parameters have been set as $d=6$, $m_g=1$, $c_{0}=1$, $c_{1}=1$, $c_{3}=-0.8$, $c_{4}=0.5$ and $q=0.5$ \\
        \textit{Critical data}: ($T_{C_{1}}=0.312557$, $P_{C_{1}}=0.177989$, $r_{c_1}=0.810774$), ($T_{C_{2}}=0.131126$, $P_{C_{2}}=0.004508$, $r_{c_2}=3.341090$) and ($T_{C_3}=0.126626$, $P_{C_3}=0.002673$, $r_{c_3}=1.933764$)}

    \label{GT_vdWtype_can}
\end{figure}

\begin{figure}[tbp]
    \epsfxsize=8cm \epsffile{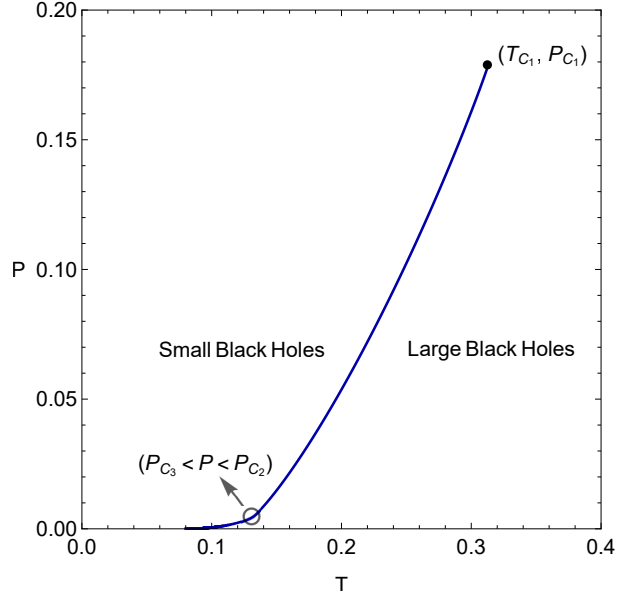}

    \caption{\textbf{vdW type phase transition in the canonical (fixed charge) ensemble:} $P-T$ diagram for a spherical black hole with [$k=+1$, $c_2=-0.1$], or a planar black hole with [$k=0$, $c_2=0.9$], or a hyperbolic black hole with [$k=-1$, $c_2=1.9$]. The other parameters have been set as $d=6$, $m_g=1$, $c_{0}=1$, $c_{1}=1$, $c_{3}=-0.8$, $c_{4}=0.5$ and $q=0.5$ \\
            \textit{Critical data}: ($T_{C_{1}}=0.312557$, $P_{C_{1}}=0.177989$, $r_{c_1}=0.810774$), ($T_{C_{2}}=0.131126$, $P_{C_{2}}=0.004508$, $r_{c_2}=3.341090$) and ($T_{C_3}=0.126626$, $P_{C_3}=0.002673$, $r_{c_3}=1.933764$)}

    \label{PT_vdWtype_can}
\end{figure}

Our investigations show that, for a certain range of parameters,
there is a lower bound for the electric charge ($Q_{b_1}$) and
there may also be an upper bound for it ($Q_{b_2}$), where for
$Q_{b_1}<Q<Q_{b_2}$, the triple point behavior can be observed.
For $Q<Q_{b_1}$ and $Q>Q_{b_2}$, one of the physical critical
points changes to an unphysical critical point which cannot
minimize the Gibbs free energy and consequently the triple point
behavior is replaced by the vdW type phase transition. Now, we
illustrate this situation for a set of TBHs in Fig.
\ref{GT_vdWtype_can}. For this purpose, we have altered the
electric charge parameter of the previous subsection (related to
Fig. \ref{GT_Triple_can}) from $q=0.8$ to $q=0.5$. Obviously, a
first order phase transition takes place for pressures in the
range $P_{C_2}<P<P_{C_1}$ and by further decreasing the pressure,
i.e., $P_{C_3}<P<P_{C_2}$, an anomaly appears in the shape of the
standard swallowtail behavior. A close-up of this anomaly is
depicted in Fig. \ref{GT_vdWtype_can}, which looks like the
reentrance of phase transition. This anomaly does not lead to any
new BH phase (or equivalently a phase transition) since it cannot
minimize the Gibbs free energy, hence we only observe the standard
SBH/LBH phase transition. This can be verified by studying the
corresponding $P-T$ diagram. In Fig. \ref{PT_vdWtype_can}, the
coexistence line of the $P-T$ diagram corresponding with the Fig.
\ref{GT_vdWtype_can} is depicted, which proves that only the
two-phase behavior exists. But, for pressures in the range
$P_{C_3}<P<P_{C_2}$, the coexistence line is curved more than any
other area. In the standard $P-T$ diagrams associated with the vdW
behavior, the coexistence curve is thoroughly smooth and uniform.

\section{Black hole chemistry in grand canonical ensemble} \label{Grand Can. BHC}

\subsection{Grand canonical partition function}
The grand partition function (${\cal Z}_{\rm{GC}}$) of the
gravitational system could be defined by a Euclidean path integral
over the tensor field $g_{\mu \nu}$ and vector field $A_\mu$ as
follows
\begin{equation}
{{\cal Z}_{\rm{GC}}} = \int {{\cal D}[g, A]} {e^{ - {{\cal I}_E}[g, A]}}\simeq e^{ - {{\cal I}_{on-shell}}(\beta,\Phi)},
\end{equation}
where $\beta$ is the inverse of Hawking temperature in terms of
the fixed potential, Eq. (\ref{temperature-grand}). In the grand
canonical ensemble, the electric charge ($Q$) fluctuates, but the
associated potential ($\Phi$) is fixed at infinity. By imposing
the fixed potential boundary condition, ${\left. {\delta {A_\nu }}
\right|_{\partial {\cal M}}} = 0$, this ensemble is established.
Thus the so-called grand potential, referred to as $G_\Phi$, can
be defined using the grand partition function. The same as before,
we will utilize the subtraction method in order to evaluate the
grand canonical on-shell action. Following the approach presented
in Sect. \ref{partition function can}, the on-shell action for the
bulk theory of TBHs is computed as
\begin{equation} \label{BH action-grand}
{{\cal I}_{BH}} = \frac{{\beta \omega _{d - 2}^k}}{{16\pi {G_d}}}\Bigg( {\Bigg[ {\frac{2}{{{\ell ^2}}}{r^{d - 1}} - m_g^2\sum\limits_{i = 1}^{d - 2} {\frac{{(i - 2)c_0^i{c_i}}}{{d - i - 1}}{r^{d - i - 1}}\prod\limits_{j = 3}^{i + 1} {{d_j}} } } \Bigg]_{{r_ + }}^{R = \infty } - 4\frac{{{d_3}}}{{{d_2}}}{\Phi ^2}r_ + ^{{d_3}}} \Bigg)
\end{equation}
in which "$R$" is an upper cutoff. In the above relation, the
fixed potential boundary condition at infinity, i.e., ${A_t}(r =
\infty ) = \Phi ({r_ + })$, is used. The on-shell action of AdS
background within the massive gravity framework (without any matter or electromagnetic field) is obtained the same as before in
Eq. (\ref{AdS action}), which is repeated below for convenience
\begin{eqnarray}
{{\cal I}_{AdS}} = \frac{{{\beta}_{0} \omega _n}}{{16\pi {G_d}}}\Bigg[ {\frac{2}{{{\ell ^2}}}{R^{d_1}} - m_g^2\sum\limits_{i = 1}^{d - 2} {\frac{{(i - 2)c_0^i{c_i}}}{{d - i - 1}}{R^{d - i - 1}}\prod\limits_{j = 3}^{i + 1} {{d_j}} } } \Bigg]. \nonumber\\
\end{eqnarray}
Demanding both the spacetimes have the same Hawking temperature
(or equivalently the same geometry) at $r=R$, i.e., ${\beta
_0}V_0(R)^{1/2} = \beta V{(R)^{1/2}}$, then subtracting the
on-shell action of the AdS background from the on-shell action of
the TBHs, one obtains
\begin{eqnarray} \label{grand canonical action}
{{\cal I}_{on - shell}} &\equiv& \mathop {\lim }\limits_{R \to
\infty } \left( {{{\cal I}_{{\rm{BH}}}} - {{\cal I}_{{\rm{AdS}}}}}
\right) \nonumber \\
&=& \frac{{\beta \omega _{n} r_ + ^{{d_3}}}}{{16\pi
{G_d}}}\Bigg[k - \frac{{r_ + ^2}}{{{\ell ^2}}} -{ 2\frac{d_3}{d_2} {\Phi ^2}}
+ m_g^2\sum\limits_{i = 1}^{d - 2} {\Big( {\frac{{(i -
1)c_0^i{c_i}}}{{r_ + ^{i - 2}}}\prod\limits_{j = 3}^i {{d_j}} }
\Big)} \Bigg].
\end{eqnarray}

\subsection{Extended phase space thermodynamics}
Working in the extended phase space, the grand canonical potential
$G_\Phi$ (sometimes is called Gibbs free energy in grand canonical
ensemble) which depends on $T$, $P$, and $\Phi$, is calculated as
\begin{eqnarray} \label{grand potential}
G_\Phi &=& {\beta ^{ - 1}}\ln {{\cal Z}_{\rm{GC}}}(T,P,\Phi ) = M- TS - \Phi Q \nonumber \\
&=& \frac{{{\omega _n}r_ + ^{{d_3}}}}{{16\pi }}\Bigg[k
- \frac{{16\pi Pr_ + ^2}}{{{d_1}{d_2}}} - 2\frac{d_3}{d_2}{\Phi ^2}+
m_g^2\sum\limits_{i = 1}^{d - 2} {\Big( {\frac{{(i -
1)c_0^i{c_i}}}{{r_ + ^{i - 2}}}\prod\limits_{j = 3}^i {{d_j}} }
\Big)} \Bigg],
\end{eqnarray}
in which we have set $G_d=1$. Moreover, it is seen that the Gibbs
free energy in the canonical ensemble is related to the grand
potential using a Legendre transform as
\begin{equation}
G_{\Phi}(T,P,\Phi)=G(T,P,Q)-\Phi Q.
\end{equation}
The above relation is a guideline to find the correct first law of
thermodynamics as $dG_\Phi=-SdT+VdP-Qd\Phi$. This can be confirmed
via the standard thermodynamic relations which will be briefly
explained. The thermodynamic variables $Q$, $V$ and $S$ are
extracted from the grand potential $G_\Phi$ as
\begin{equation}
Q =  - {\left( {\frac{{\partial {G_\Phi }}}{{\partial \Phi }}} \right)_{T,P}} = \frac{{{\omega _n}}}{{4\pi }}q,
\end{equation}
\begin{equation}
V = {\left( {\frac{{\partial {G_\Phi }}}{{\partial P}}} \right)_{T,\Phi }} = \frac{{{\omega _{{d_2}}}}}{{{d_1}}}r_ + ^{{d_1}},
\end{equation}
and
\begin{equation} \label{entropy-grand}
S =  - {\left( {\frac{{\partial {G_\Phi }}}{{\partial T}}} \right)_{P,\Phi }} = \frac{{{\omega _n}}}{4}r_ + ^{{d_2}}.
\end{equation}
The above thermodynamic variables are in full agreement with the
previous results, i.e. Eqs. (\ref{charge}), (\ref{volume}) and
(\ref{entropy}), that we have obtained before. Therefore, we
deduce that all intensive and corresponding extensive variables
satisfy the first law of black hole thermodynamics in the grand
potential representation. The extended Smarr formula does not
depend on the ensemble one is dealing with, thus the same relation
as before in Eq. (\ref{Smarr}) is obtained again. Moreover,
according to Sect. \ref{Can Thermodynamics}, if one considers the
massive coupling constants as the new thermodynamic variables, the
first law in the grand potential representation is generalized as
$dG_\Phi=-SdT+VdP-Qd\Phi+\sum\limits_{i = 1}^{d - 2} {{{\cal
C}_i}d{c_i}}$.

\subsection{Holographic phase transitions}

The grand canonical equation of state is obtained as
\begin{equation} \label{Grand EOS}
P = \frac{{{d_2}\tilde T}}{{4{r_ + }}} - \frac{{{d_2}{d_3} k_{\rm{eff}}^{\rm{(GC)}}}}{{16\pi r_ + ^2}}- \frac{{m_g^2}}{{16\pi }}\sum\limits_{i = 3}^{d - 2} {\Big( {\frac{{c_0^i{c_i}}}{{\,r_ + ^i}}\prod\limits_{j = 2}^{i + 1} {{d_j}} } \Big)},
\end{equation}
where, in this ensemble, the effective topological factor
$k_{\rm{eff}}^{\rm{(GC)}}$ and the shifted Hawking temperature
$\tilde T $ are given by
\begin{equation}
k_{\rm{eff}}^{\rm{(GC)}} \equiv [k + m_g^2c_0^2{c_2} - 2({d_3}/{d_2}){\Phi ^2}]
\end{equation}
and
\begin{eqnarray}
\tilde T &=& T - \frac{{m_g^2{c_0}{c_1}}}{{4\pi }} =
\frac{{{d_2}{d_3}k - 2\Lambda r_ + ^2 - 2{{({d_3}\Phi )}^2} +
m_g^2\sum\limits_{i = 2}^{d - 2} {\Big( {\frac{{c_0^i{c_i}}}{{r_ +
^{i - 2}}}\prod\limits_{j = 2}^{i + 1} {{d_j}} } \Big)} }}{{4\pi
{d_2}{r_ + }}}.
\end{eqnarray}
Obviously, in the Einstein limit, i.e. $m_{g} \to 0$, we cannot
observe any critical behavior, which means that there does not
exist criticality in the grand canonical ensemble of AdS black
holes in GR with or without the linear electromagnetic Maxwell
fields.

The critical point occurs at the spike like divergence of specific
heat at constant pressure i.e., an inflection point in the $P-v$
(or equivalently $P-r_+$) diagram and can be found by using Eq.
\ref{inflection points} which for the grand canonical equation of
state (\ref{Grand EOS}) leads to
\begin{equation} \label{CPE-GrandCan}
2 k_{\rm{eff}}^{\rm{(GC)}} r_ + ^{d_4}
+ m_g^2\sum\limits_{i = 3}^{d - 2} {\Big( {i(i - 1)c_0^i{c_i}r_ + ^{d - i - 2}\prod\limits_{j = 4}^{i + 1} {{d_j}} } \Big)}  = 0.
\end{equation}
Now, to be more specific, following Sect. \ref{Canonical BHC},
we analyze the equations of state and holographic phase
transitions for case by case of TBHs with detail. As before, we
generally concentrate on the isobaric curves of $G-T$ diagrams
since all the essential information about the critical behaviors
can be extracted from them. In addition, Eqs. (\ref{Grand EOS})
and (\ref{CPE-GrandCan}) help us to find that the combination
$k_{\rm{eff}}^{\rm{(C)}} \equiv [k + m_g^2c_0^2{c_2}]$ in the
canonical ensemble is replaced by the $k_{\rm{eff}}^{\rm{(GC)}}
\equiv [k + m_g^2c_0^2{c_2} - 2({d_3}/{d_2}){\Phi ^2}]$ in the
grand canonical ensemble. So the same critical behavior with the
same critical points would be found for the case of the spherical,
planar, and hyperbolic black holes if the same value for
$k_{\rm{eff}}^{\rm{(GC)}}$ are provided. In this view, TBHs in
massive gravity at their critical point may be indistinguishable.

\subsubsection{van der Waals (vdW) phase transition}

In order to observe the vdW behavior in a given black hole
spacetime, one physical critical point must exist in the
thermodynamic phase space which minimize the Gibbs free energy.
Within the framework of the grand canonical ensemble, this can be
obtained in the spacetime dimensions with the range $d \ge
5$.\footnote{Comparing with the vdW critical behavior in the
canonical ensemble, it is seen that, in $d=4$, only canonical vdW
phase transition can take place.} In $d=5$, only the first three
massive couplings ($c_1$, $c_2$ and $c_3$) appears. Regarding Eq.
\ref{CPE-GrandCan} and following the approach presented in Sect.
\ref{Canonical BHC}, the critical point of the massive charged
TBHs can be obtained from the root of following relation
\begin{equation} \label{vdW-CPE-GC}
k_{\rm{eff}}^{\rm{(GC)}} {r_ + } + 3{d_4}m_g^2c_0^3{c_3} = 0,
\end{equation}
in which we have assumed that, in higher dimensions, the other
massive couplings $c_i$ ($i \ge 4$) vanish. This is the simplest
way to find the vdW behavior in arbitrary dimensions, and, of
course, it is permissible to consider higher order graviton
self-interaction terms and then observe a vdW phase transition by
using a fine tuning of massive couplings. Considering Eq.
(\ref{vdW-CPE-GC}), the critical radius is easily obtained as
\begin{equation} \label{vdW-CP-Grand}
{r_c} = \frac{{3{d_4}m_g^2c_0^3{c_3}}}{k_{\rm{eff}}^{\rm{(GC)}}},
\end{equation}
with the following constraints on the parameters
\begin{equation} \label{vdW-CP-first constraint}
{c_3} > 0 \leftrightarrow {\Phi ^2} < \frac{{{d_2}[k + m_g^2c_0^2{c_2}]}}{{2{d_3}}},
\end{equation}
and
\begin{equation} \label{vdW-CP-second constraint}
{c_3} < 0 \leftrightarrow {\Phi ^2} > \frac{{{d_2}[k + m_g^2c_0^2{c_2}]}}{{2{d_3}}}.
\end{equation}
As seen, there are two strict limitations on the value of the
$U(1)$ potential, $\Phi$. According to these constraints, one has
to assume $[k + m_g^2c_0^2{c_2}]>0$ when ${c_3}$ is positive
definite, while there is no such constraint when ${c_3}$ is
negative. As we will see in a moment, the first constraint
(\ref{vdW-CP-first constraint}) does not lead to vdW phase
transition since the associated pressure and temperature are
negative definite. The thermodynamic pressure and temperature at
the critical point (\ref{vdW-CP-Grand}) are given as
\begin{equation} \label{critical pressure-GC}
{P_C} =  - \frac{{{d_2}{d_3}{d_4}m_g^2c_0^3{c_3}}}{{16\pi r_C^3}},
\end{equation}
and
\begin{equation} \label{critical temp-GC}
{\tilde T_C} =  - \frac{{3{d_3}{d_4}m_g^2c_0^3{c_3}}}{{4\pi r_C^2}}.
\end{equation}
According to the above relations, the massive coupling $c_3$ must
be negative definite, so the condition (\ref{vdW-CP-first
constraint}) does not lead to criticality at all.

\begin{figure}[tbp]
    \epsfxsize=7cm \epsffile{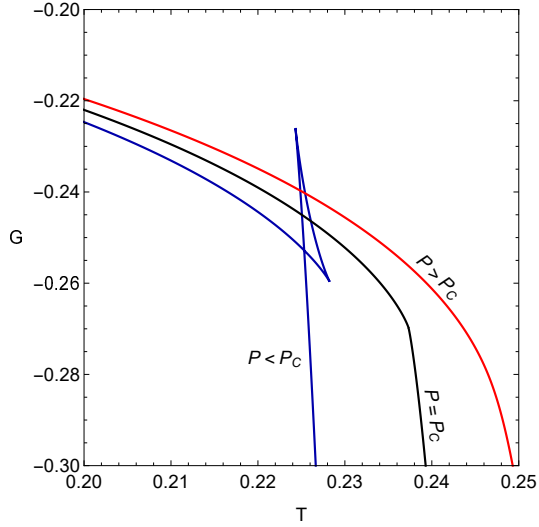}

    \caption{\textbf{vdW phase transition in the grand canonical (fixed potential) ensemble:} the $G-T$ diagram for a spherical black hole with [$k=+1$, $c_2=2$], or a planar black hole with [$k=0$, $c_2=3$], or a hyperbolic black hole with [$k=-1$, $c_2=4$]. The other parameters have been set as $d=5$, $m_g=1$, $c_{0}=1$, $c_{1}=1$, $c_{3}=-3$ and $\Phi=0.1$ \\
    \textit{Critical data}: ($T_C=0.237321$, $P_C=0.013086$)}

    \label{GT_vdW_grand}
\end{figure}

In Fig. \ref{GT_vdW_grand}, according to Eqs.
(\ref{vdW-CP-Grand}), (\ref{vdW-CP-first constraint}) and
(\ref{vdW-CP-second constraint}), the massive coupling
coefficients and $U(1)$ potential $\Phi$ have been adjusted in a
way to produce a vdW behavior. As the canonical ensemble case, the
swallowtail behavior is observed for pressures in the range
$P<P_C$ which indicates the existence of two-phase behavior. As
seen, the single phase behavior takes place for pressures in the
range $P>P_C$.

The obtained critical data, Eqs. (\ref{vdW-CP-Grand}),
(\ref{critical pressure-GC}) and (\ref{critical temp-GC}), satisfy
the grand canonical universal ratio as
\begin{equation}  \label{ratio-grand-shifted}
\frac{{{P_C}{r_c}}}{{{{\tilde T}_C}}} = \frac{{d - 2}}{{12}} \Longleftrightarrow \frac{{{P_C}{v_c}}}{{{{\tilde T}_C}}} = \frac{1}{3} \,.
\end{equation}
Interestingly, the universal ratio at critical point when is
written down in terms of the critical specific volume ($v_c$) and
the shifted Hawking temperature (${\tilde T}_C$), i.e.,
${P_C}{v_c}/{{\tilde T}_C}$, is constant and does not depend on
the spacetime dimensions ($d$). To our knowledge, till now, this
case and the case of charged-AdS BHs in the PMI-Einstein gravity
(see Ref. \cite{Vahidinia2013}) are the only examples of such a BH
spacetime with a constant universal ratio. At this stage, the
standard universal ratio may also be written as
\begin{equation} \label{ratio-grand}
\frac{{{P_C}{v_c}}}{{{T_C}}} = \frac{{{d_3}{d_4}c_0^3{c_3}}}{{3{d_3}{d_4}c_0^3{c_3} - {c_0}{c_1}r_c^2}}
\end{equation}
Expanding the above universal ratio around the small values of graviton mass yields
\begin{equation} \label{expansion-ratio-GC}
\frac{{{P_C}{v_c}}}{{{T_C}}} ={ \frac{1}{3} + \frac{m_g^4{c_0^4{c_1}{c_3}d_2^2{d_4}}}{{{d_3}{{\left( {{d_2}k - 2{d_3}{\Phi ^2}} \right)}^2}}} + O(m_g^6)}.
\end{equation}
But, it does not mean that the massless limit of massive gravity
in the grand canonical ensemble leads to the outcome of Einstein
gravity. In fact, according to Eq. (\ref{vdW-CP-Grand}), the
critical radius vanishes in the massless limit ($m_g=0$), as
expected, since there exists no grand canonical criticality for
TBHs in Einstein gravity. So the massless limit ($m_g=0$) of Eq.
(\ref{expansion-ratio-GC}) does not appear at all since there is
not any critical point at this limit.

\subsubsection{Reentrant phase transition (RPT)}

As stated in \cite{HendiDehghani2019}, in order to have a RPT
phenomenon, the critical point equation must admit two positive
critical radii in which the associated pressures and temperatures
are positive definite, while only one of the critical points,
referred to as ($T_C$, $P_C$), can minimize the Gibbs free energy.
Simply, a real and inhomogeneous polynomial equation of
second-degree of $r_+$ can produce two critical radii. This is
permissible in $d \ge 6$ dimensions which implies the first four
massive couplings ($c_1$, $c_2$, $c_3$ and $c_4$) have to be
nonzero. Assuming that these massive couplings ($c_1$, $c_2$,
$c_3$ and $c_4$) are nonzero and the other couplings vanish in
higher dimensions, one gets
\begin{equation}
k_{\rm{eff}}^{\rm{(GC)}} r_ + ^2 + 3{d_4}m_g^2c_0^3{c_3}{r_ + } +6{d_4}{d_5}m_g^2c_0^4{c_4} = 0.
\end{equation}
The above equation of critical point can be solved simply as
\begin{eqnarray}
&&{r_{c_1}},{r_{c_2}} = \frac{{ - 3{d_4}{m^2}c_0^3{c_3} \pm \sqrt \Delta  }}{{2 k_{\rm{eff}}^{\rm{(GC)}}}},\quad \Delta>0, \nonumber  \\
&&\Delta = - 24 k_{\rm{eff}}^{\rm{(GC)}} ({d_4}{d_5}{m^2}c_0^4{c_4})+{(3{d_4}{m^2}c_0^3{c_3})^2}.
\end{eqnarray}
It is obvious that the above relation can predict one or at most
two positive critical radii for the equation of state of TBHs.
Since the reentrant behavior of phase transition takes place
whenever the critical point equation possesses two positive roots,
so looking for this case, the following conditions should be
satisfied
\begin{equation}
{r_{{c_1}}} + {r_{{c_2}}} = \frac{{ - 3{d_4}{m^2}c_0^3{c_3}}}{k_{\rm{eff}}^{\rm{(GC)}}} > 0, \quad {r_{{c_1}}}{r_{{c_2}}} = \frac{{6{d_4}{d_5}{m^2}c_0^4{c_4}}}{{k_{\rm{eff}}^{\rm{(GC)}}}} > 0.
\end{equation}
According to the above constraints, when the effective topological
factor $k_{\rm{eff}}^{\rm{(GC)}}$ is positive definite, two
critical points can be found assuming that $c_{3}<0$ and
$c_{4}>0$, and when  $k_{\rm{eff}}^{\rm{(GC)}}<0$, one has to
assume  $c_{3}>0$ and  $c_{4}<0$.

\begin{figure}[tbp]
    \epsfxsize=8cm \epsffile{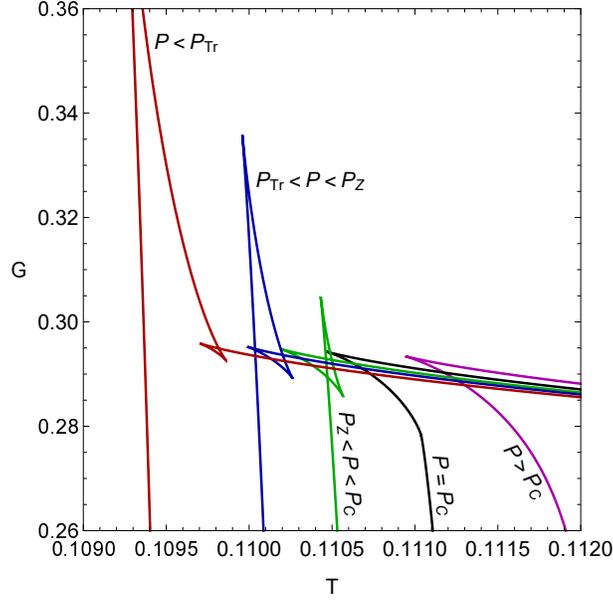}

    \caption{\textbf{RPT in the grand canonical (fixed potential) ensemble:} the $G-T$ diagram for a spherical black hole with [$k=+1$, $c_2=-0.2$], or a planar black hole with [$k=0$, $c_2=0.8$], or a hyperbolic black hole with [$k=-1$, $c_2=0.2$]. The other parameters have been set as $d=6$, $m_g=1$, $c_{0}=1$, $c_{1}=1$, $c_{3}=-1$, $c_{4}=0.9$ and $\Phi=0.1$ \\
        \textit{Critical data}: ($T_C=0.237321$, $P_C=0.013086$), ($T_Z=0.1100131$, $P_Z=0.0017463$) and ($T_{Tr}=0.1099575$, $P_{Tr}=0.001724$)}

    \label{GT_RPT_grand}
\end{figure}

Now, using the obtained information, we illustrate a typical
example of the RPT phenomenon in the grand canonical ensemble. In
Fig. \ref{GT_RPT_grand}, the $G-T$ diagrams for a set of charged
TBHs are depicted. As seen, two new critical points, referred to
as ($T_Z$, $P_Z$) and ($T_{Tr}$, $P_{Tr}$), emerge in the
thermodynamic phase space. For pressures in the range $P_Z<P<P_C$,
a first order phase transition occurs as temperature decreases.
For $P_{Tr}<P<P_Z$, as temperature decreases monotonically, a
first-order phase transition is initially observed, and then, a
finite jump (discontinuity) appears in the global minimum of the
Gibbs free energy, which displays the zeroth-order phase
transition. This behavior is exactly the standard RPT in the
subject of black hole chemistry seen in many black hole spacetimes
before. This phenomenon reminds us of those critical behaviors
present in multicomponent fluids and liquid crystals
\cite{NarayananKumar1994}. In addition, comparing with Ref.
\cite{DehghaniHendiMann2019}, this phenomenon occurs qualitatively
in the same dimensions as neutral TBHs.

\subsubsection{Triple point and small/intermediate/large black hole (SBH/IBH/LBH) phase transition} \label{triple-grand}

\begin{figure}[tbp]
    \epsfxsize=8.5cm \epsffile{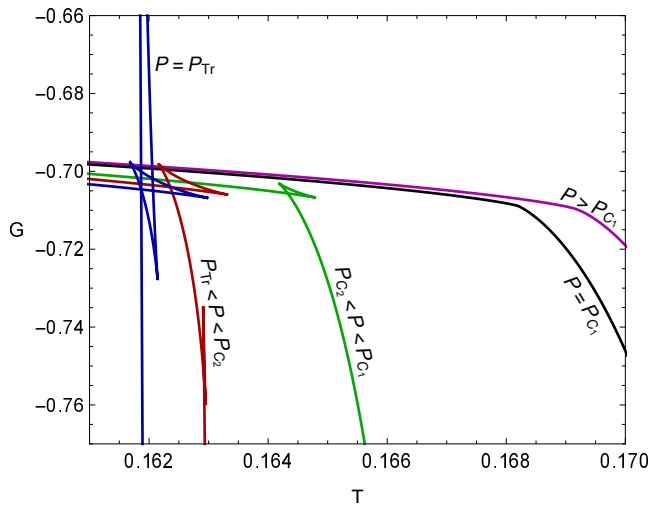}

    \caption{\textbf{Triple point in the grand canonical (fixed potential) ensemble:} the $G-T$ diagram for a spherical black hole with [$k=+1$, $c_2=0.8$], or a planar black hole with [$k=0$, $c_2=1.8$], or a hyperbolic black hole with [$k=-1$, $c_2=2.8$]. The other parameters have been set as $d=7$, $m_g=1$, $c_{0}=1$, $c_{1}=1$, $c_{3}=-1.8$, $c_{4}=1.3$, $c_{5}=-0.7$ and $\Phi=0.2$ \\
        \textit{Critical data}: ($T_{C_{1}}=0.168188$, $P_{C_{1}}=0.009558$), ($T_{C_{2}}=0.1632424$, $P_{C_{2}}=0.604562$) and ($T_{Tr}=0.161867$, $P_{Tr}=0.570678$)}

    \label{GT_Triple_grand}
\end{figure}

In the grand canonical ensemble, the analogue of triple point may
be found whenever the TBH equation of state is supplemented by
higher order interacting potentials of massive gravitons up to the
fifth interaction terms. Hence, assuming that the only first five
massive couplings are nonzero, equation of critical point
(\ref{Grand EOS}) reduces to the following polynomial
\begin{eqnarray}
&&k_{\rm{eff}}^{\rm{(GC)}} r_ + ^3 + 3{d_4}m_g^2c_0^3{c_3}r_ + ^2
+ 6{d_4}{d_5}m_g^2c_0^4{c_4}{r_ + }   +
10{d_4}{d_5}{d_6}m_g^2c_0^5{c_5} = 0.
\end{eqnarray}
Investigation of the exact solutions of three critical points is
not possible analytically, and so we apply the numerical
techniques. To do that, we have suitably tuned the massive
couplings to produce three critical points, in which two of them
are physical and one of them cannot minimize the Gibbs free
energy. The corresponding critical behavior via the $G-T$ diagram
is depicted in Fig. \ref{GT_Triple_grand}. Obviously, a critical
triple point ($T_{Tr}$, $P_{Tr}$) emerges, and, consequently,
three-phase behavior appears. Qualitatively, this critical
behavior is the same as its counterpart in the canonical ensemble.
But, in $d=6$ dimensions, the triple point behavior can solely
take place in the canonical ensemble. In addition, this phenomenon
occurs qualitatively in the same dimensions as neutral TBHs.

\subsubsection{vdW type phase transition} \label{vdW-type Grand}
\begin{figure}[tbp]
	\epsfxsize=8cm \epsffile{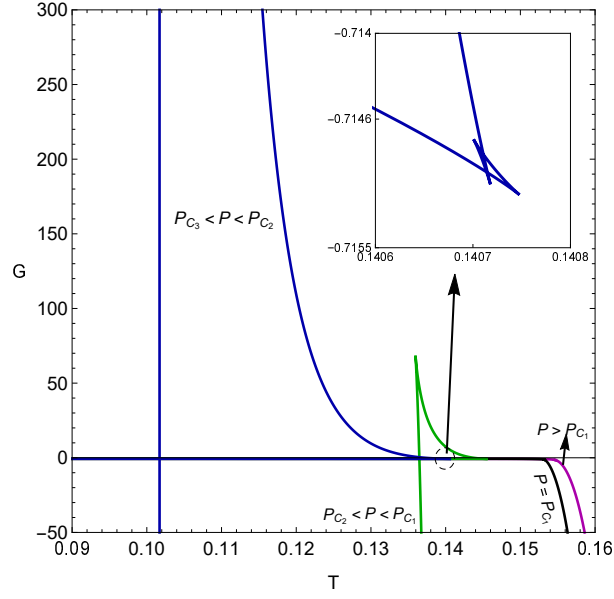}
	
	\caption{\textbf{vdW type phase transition in the grand canonical (fixed potential) ensemble:} the $G-T$ diagram for a spherical black hole with [$k=+1$, $c_2=0.8$], or a planar black hole with [$k=0$, $c_2=1.8$], or a hyperbolic black hole with [$k=-1$, $c_2=2.8$]. The other parameters have been set as $d=7$, $m_g=1$, $c_{0}=1$, $c_{1}=1$, $c_{3}=-1.8$, $c_{4}=1.3$, $c_{5}=-0.7$ and $\Phi=0.3$ \\
		\textit{Critical data}: ($T_{C_{1}}=0.153181$, $P_{C_{1}}=0.0048008$, $r_{c_1}=5.45278$), ($T_{C_{2}}=0.141182$, $P_{C_{2}}=0.00060427$, $r_{c_2}=1.97626$) and ($T_{C_3}=0.140511$, $P_{C_3}=0.00021291$, $r_{c_3}=2.35357$)}
	
	\label{GT_vdWtype_grand}
\end{figure}

\begin{figure}[tbp]
	\epsfxsize=8cm \epsffile{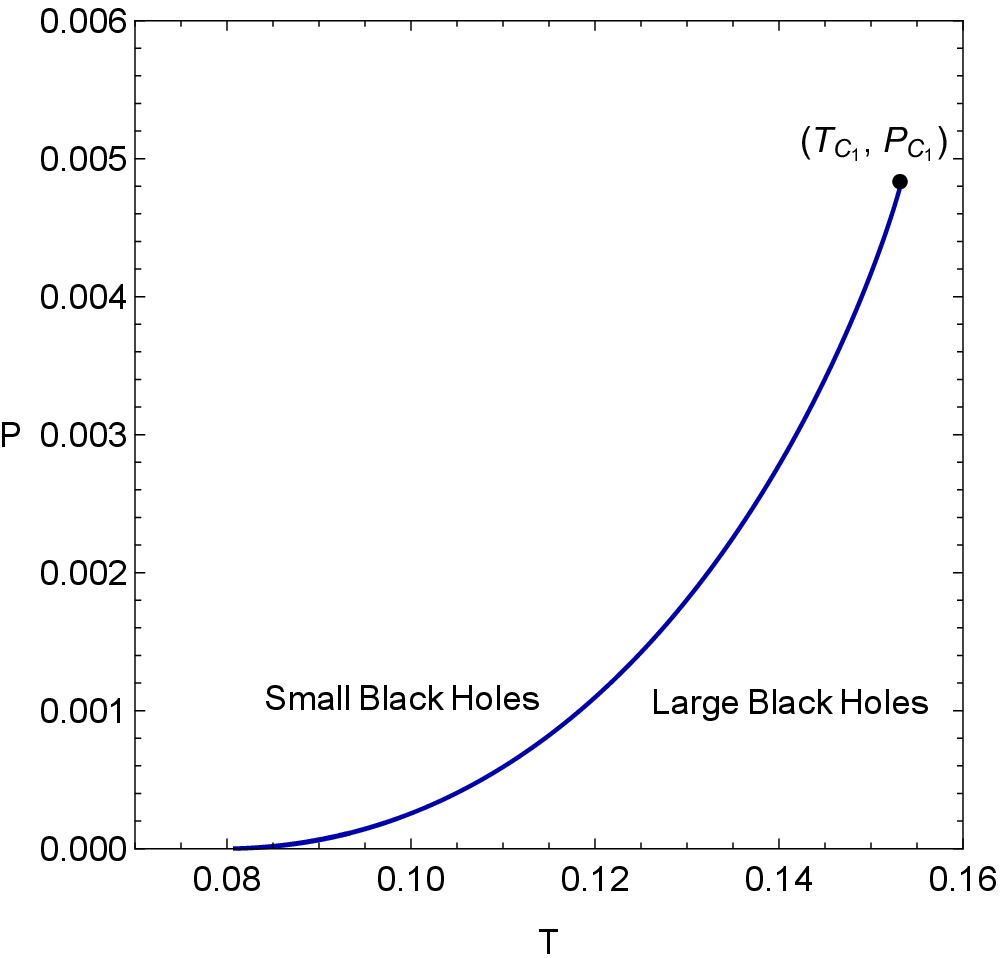}
	
	\caption{\textbf{vdW type phase transition in the grand canonical (fixed potential) ensemble:} the $P-T$ diagram for a spherical black hole with [$k=+1$, $c_2=0.8$], or a planar black hole with [$k=0$, $c_2=1.8$], or a hyperbolic black hole with [$k=-1$, $c_2=2.8$]. The other parameters have been set as $d=7$, $m_g=1$, $c_{0}=1$, $c_{1}=1$, $c_{3}=-1.8$, $c_{4}=1.3$, $c_{5}=-0.7$ and $\Phi=0.3$ \\
		\textit{Critical data}: ($T_{C_{1}}=0.153181$, $P_{C_{1}}=0.0048008$, $r_{c_1}=5.45278$), ($T_{C_{2}}=0.141182$, $P_{C_{2}}=0.00060427$, $r_{c_2}=1.97626$) and ($T_{C_3}=0.140511$, $P_{C_3}=0.00021291$, $r_{c_3}=2.35357$)}
	
	\label{PT_vdWtype_grand}
\end{figure}

Here in parallel with Sect. \ref{vdW-type PT}, we discuss the vdw type phase transition in the grand canonical ensemble. For this phenomenon, the essential requirement is the existence of three (possible) critical points for the equation of state (\ref{Grand EOS}) in which only one of them minimizes the Gibbs free energy. Elementally, this can happen by varying the parameter space of theory in the spacetime dimensions which triple point phenomenon takes place (since for the case of triple point, the equation of state admits three possible critical points). So one can draw a conclusion that the vdW type behaviour and triple point phenomenon always show up in the same spacetime dimensions. To see this, we can alter the
electric potential parameter ($\Phi$) of the previous example in  Sect. \ref{triple-grand} (related to Fig. \ref{GT_Triple_grand}) from $\Phi=0.2$ to  $\Phi=0.3$. In this case, the critical point equation (\ref{CPE-GrandCan}) still admits three (positive) critical radii in which the associated temperatures and pressures are positive definite. However, only one of the critical points, referred to as ($T_{C_{1}}$, $P_{C_{1}}$), is physical since it is the only critical point that minimizes the Gibbs free energy. The corresponding critical phenomenon can be understood using the $G-T$ and $P-T$ diagrams illustrated, respectively, in Figs. \ref{GT_vdWtype_grand} and \ref{PT_vdWtype_grand} explicitly confirms the vdW type phase transition in the grand canonical ensemble. This phenomenon persists in higher dimensions ($d \ge 7$) as well.

\section{Critical exponents} \label{critical exponents}

Critical exponents ($\alpha$, $\beta$, $\gamma$ and $\delta$)
determine the behavior of thermodynamic quantities in the
neighborhood of critical points, so various critical exponents
imply different behaviors in the phase diagrams. So far, in the
realm of Statistical Mechanics, it has been confirmed that these
exponents do not depend on the microscopic details of a physical
system, and they are highly affected by spatial dimensions,
symmetries and the range of interactions
\cite{Huang2001Book,Zinn-Justin2007}. Let us now compute them in
the subject of charged black hole chemistry in massive gravity
following the approach of Refs.
\cite{KubiznakMann2012,Gunasekaran2012Mann}.

In order to obtain the critical exponents, hereafter, we work with
the thermodynamic quantities in terms of the following variables
\begin{equation}
p \equiv \frac{P}{{{P_C}}}, \quad\tau  \equiv \frac{{T -
{T_C}}}{{{T_C}}}, \quad w \equiv \frac{{v - {v_c}}}{{{v_c}}},
\quad \rho_c \equiv \frac{P_C v_c}{T_C}.
\end{equation}
The exponent $\alpha$ specifies the behavior of the specific heat
at constant volume as
\begin{equation}
{C_v} = T{\left( {\frac{{\partial S}}{{\partial T}}} \right)_v}
\propto {\left| \tau \right|^{ - \alpha }}.
\end{equation}
Using the relations of the entropy in different ensembles,  Eqs.
(\ref{entropy}) and (\ref{entropy-grand}), it is inferred that
$C_v=0$ since the entropy does not depend on $T$, and so, we have
$\alpha=0$ in both the canonical and grand canonical ensembles.

The exponent $\beta$ determines the behavior of the order
parameter $\eta$ on the isotherms as
\begin{equation} \label{exponent beta}
\eta = v_l - v_g \propto |\tau|^{\beta}.
\end{equation}
To compute this exponent, first, we expand the equation of states
in the canonical and grand canonical ensembles, Eqs. (\ref{Can
EOS}) and (\ref{Grand EOS}), near the critical point
\begin{equation} \label{EOS expansion}
p = 1 + \frac{\tau}{{{\rho _c}}}(1 - w) + h(v_c,{c_i},q){w^3} + O(t{w^2},{w^4}),
\end{equation}
where the function $h(v_c,{c_i},q)$ takes different forms in
different ensembles. This function in the canonical ensemble is
given by \footnote{In the published version, there is a minor typo in the coefficients of the electromagnetic part of this function (only for the case of canonical ensemble). Since the details of this function is not important at all, the final result is the same as before.}
\begin{eqnarray}
h \equiv \frac{{m_g^2{c_0}{c_1} - 4\pi {T_C}}}{{4\pi {P_C}{v_c}}} - \frac{{{4^{2d}}(2d - 3){d_1}{q^2}}}{{3 \times {2^{10}}\pi {P_C}{d^{(2d - 5)}}v_c^{2{d_2}}}} + \frac{{4{d_3}k_{{\rm{eff}}}^{{\rm{(C)}}}}}{{{d_2}\pi {P_C}v_c^2}} + \frac{{40m_g^2c_0^3{c_3}{d_3}{d_4}}}{{d_2^2\pi {P_C}v_c^3}} + \frac{{320m_g^2c_0^4{c_4}{d_3}{d_4}{d_5}}}{{d_2^3\pi {P_C}v_c^4}} + O\left( {\frac{1}{{v_c^5}}} \right)
\end{eqnarray}
while in the grand canonical ensemble it reads as
\begin{eqnarray}
h &\equiv& \frac{{m_g^2{c_0}{c_1} - 4\pi {T_C}}}{{4\pi
{P_C}{v_c}}} + \frac{{4{d_3} k_{\rm{eff}}^{\rm{(GC)}}}}{{{d_2}\pi
{P_C}v_c^2}}+ \frac{{40m_g^2c_0^3{c_3}{d_3}{d_4}}}{{d_2^2\pi
{P_C}v_c^3}} +\frac{{320m_g^2c_0^4{c_4}{d_3}{d_4}{d_5}}}{{d_2^3\pi
{P_C}v_c^4}} + O\left( {\frac{1}{{v_c^5}}} \right).
\end{eqnarray}
As will be clear, the final results do not depend on the function
$h(v_c,c_i,q)$ and so the details of this function is not
important at all. Differentiating the expansion (\ref{EOS
expansion}) for a fixed $\tau<0$ yields
\begin{equation}
dp = \Big( { - \frac{\tau}{{{\rho _c}}} + 3h(v_c,{c_i},q){w^2}} \Big)dw.
\end{equation}
Using the above relation and Maxwell's equal area law ($\oint {vdP = 0}$), one finds
\begin{equation} \label{Maxwell's construction}
\int_{{w_l}}^{{w_s}} {wdp = } \frac{{ - \tau}}{{2{\rho _c}}}(w_s^2 - w_l^2) + \frac{{3h}}{4}(w_s^4 - w_l^4) = 0,
\end{equation}
where ${w_l}$ and ${w_s}$ denote the volume of large and small
black holes, respectively. Now, using Eq. (\ref{EOS expansion})
and the fact that the pressure of different black hole phases
keeps unchanged at the critical point, we get
\begin{equation} \label{pressureATtransition}
1 + \frac{\tau}{{{\rho _c}}}(1 - {w_s}) + h(v_c,{c_i},q)w_s^3 = 1 + \frac{\tau}{{{\rho _c}}}(1 - {w_l}) + h(v_c,{c_i},q)w_l^3.
\end{equation}
Equations (\ref{Maxwell's construction}) and
(\ref{pressureATtransition}) admit a unique nontrivial solution as
${w_s} =  - {w_l} = \sqrt {\frac{{ - \tau}}{{{\rho _c}h}}}$.
Therefore, the behavior of the order parameter $\eta$ is obtained
simply as
\begin{equation}
\eta = v_l - v_s = v_c (w_l - w_s) = 2 v_c  \sqrt {\frac{{ - \tau}}{{{\rho _c}h}}} \propto |\tau|^{1/2},
\end{equation}
which, according to Eq. (\ref{exponent beta}), confirms that $\beta=1/2$ for both the canonical and grand canonical ensembles.

The exponent $\gamma$ is extracted from the definition of the isothermal compressibility near the critical point given by
\begin{equation} \label{exponent gamma}
{\kappa _T} =  - \frac{1}{v}{\left. {\frac{{\partial v}}{{\partial P}}} \right|_T} \propto {\left| \tau  \right|^{ - \gamma }}.
\end{equation}
Considering both the ensembles, we differentiate the expansion of equations of state (\ref{EOS expansion}) to get
\begin{equation}
{\left. {\frac{{\partial P}}{{\partial v}}} \right|_T} = \frac{{ - {P_c} \tau}}{{{\rho _c}{v_c}}}  + O(tw,{w^2}),
\end{equation}
and then, by using ${\left. {\frac{{\partial v}}{{\partial P}}} \right|_T} = {\left( {{{\left. {\frac{{\partial P}}{{\partial v}}} \right|}_T}} \right)^{ - 1}}$, we obtain
\begin{equation}
{\kappa _T} =  - \frac{1}{v}{\left. {\frac{{\partial v}}{{\partial P}}} \right|_T} \propto \frac{{{\rho _c}{v_c}}}{{{P_C}}}\frac{1}{\tau }.
\end{equation}
According to Eq. (\ref{exponent gamma}), this relation confirms that $\gamma  = 1$ for both the canonical and grand canonical ensembles.

The exponent $\delta$, which specifies the shape of the critical isotherm, is defined as
\begin{equation} \label{exponent delta}
\left| {P - {P_c}} \right| \propto {\left| {v - {v_c}} \right|^\delta }.
\end{equation}
This exponent obtains easily by putting $\tau=0$ in the expansion (\ref{EOS expansion}), which leads to
\begin{equation}
P - {P_C} = \frac{{{P_C}h(v_c,{c_i},q)}}{{v_c^3}}{(v - {v_c})^3}.
\end{equation}
Comparing with Eq. (\ref{exponent delta}), we find $\delta=3$ for both the canonical and grand canonical ensembles.

To sum up, the critical exponents are obtained as $\alpha=0$,
$\beta=1/2$, $\gamma=1$ and $\delta=3$ for both the canonical and
grand canonical ensembles, the same as van der Waals fluid.  In
addition, the same results for the critical exponents in the
(grand) canonical ensemble would be obtained if one uses the
definition of $w$ in terms of thermodynamic volume $V$ as $w
\equiv \frac{{V - {V_C}}}{{{V_C}}}$.

\section{Conclusion} \label{Conclusion}

In the context of gauge/gravity duality, AdS BH solutions of dRGT
massive gravity are dual to homogeneous and isotropic condensed
matter systems with broken translational invariance
\cite{Vegh2013,BlakeTong2013,Davison2013}. More importantly, as
indicated in a series of papers
\cite{Vegh2013,Alberte2016,Alberte2016SolidHolography,Alberte2018HolographicPhonons,Alberte2018BHelasticity},
AdS BH solutions in massive gravity theories can effectively
describe certain properties of different types (phases) of matter
(solids, liquids etc) such as elasticity. Motivated by these
facts, we extensively explored the chemistry of charged BH
solutions of dRGT massive gravity with a suitable degenerate
reference metric in the extended phase space and remarkably found
a range of novel phase transitions in various ensembles close to
realistic ones in real world.

To be more specific, we introduced the $U(1)$ charged TBHs in
arbitrary dimensions by considering the full nonlinear theory of
dRGT massive gravity with all the higher order graviton
self-interactions. We evaluated the renormalized on-shell action
in both the canonical (fixed charge) and grand canonical (fixed
potential) ensembles with appropriate boundary conditions to
obtain the corresponding semi-classical partition functions. By
extracting the thermodynamic quantities from the partition
functions in both the canonical and grand canonical ensembles, we
have shown these quantities satisfy the extended first law of
thermodynamics in different representations. In addition, the
validity of the Smarr formula in the extended phase space has been
checked for this class of charged TBHs.

Next, since critical behaviors and nature of possible phase
transition(s) are crucially dependent on the specific choice of
ensemble, we focused on holographic phase transitions in various
ensembles. In this regard, we explicitly demonstrated that vdW
phase transition (in $d \ge 4$), vdW type phase transition (in $d
\ge 6$) and the SBH/IBH/LBH phase transition associated with the
triple point (in $d \ge 6$) are present in the canonical ensemble.
Here, the absence of the phenomenon of RPT in this ensemble is
interesting (we proved this claim in $d=5,6$ analytically, and, by
using numerical investigation in $d=7,8,9$ dimensions, this
phenomenon did not find too. Whether or not such
phenomenon exists for higher dimensional TBHs ($d \ge 7$) in the canonical ensemble remains an open question.). In the case of the grand canonical
ensemble, we observe the vdW critical behavior (in $d \ge 5$), RPT (in $d \ge 6$), vdW type phase transition (in $d \ge 7$) and triple point (in $d \ge 7$) in contrast to
Einstein gravity which only phase transition takes place in the
canonical ensemble of charged or rotating BHs. So, these critical
phenomena may commence appearing in diverse dimensions depending
on the ensemble one is dealing with. Here for convenience, we have
summarized the final results in table \ref{tab:ensembles}.

\begin{table}[]
	\caption{Holographic phase transitions in canonical (fixed charge) and grand canonical (fixed potential) ensembles}
	\label{tab:ensembles}
	\begin{tabular}{|c|c|c|ll}
		\cline{1-3}
		\textbf{Phase transition (PT)} & \textbf{Canonical ensemble} & \textbf{Grand Canonical ensemble} &  &  \\ \cline{1-3}
		\textbf{vdW PT} & in $d \ge 4$ & in $d \ge 5$ &  &  \\ \cline{1-3}
		\textbf{RPT} & not seen & in  $d \ge 6$ &  &  \\ \cline{1-3}
		\textbf{Triple Point} & in $d \ge 6$ & in  $d \ge 7$ &  &  \\ \cline{1-3}
		\textbf{vdW type PT} & in $d \ge 6$ & in $d \ge 7$  &  &  \\ \cline{1-3}
	\end{tabular}
\end{table}

These results show that, within the framework of massive
gravity, critical behavior and phase transition(s) of charged-TBHs
in the grand canonical ensemble are qualitatively the same as the
uncharged (neutral) black holes. In fact, in the case of the grand
canonical ensemble, the scalar potential ($\Phi$) is absorbed into
the effective topological factor $k_{\rm{eff}}^{\rm{(GC)}} \equiv
[k + {m^2}c_0^2{c_2} - 2({d_3}/{d_2}){\Phi ^2}]$, and thus, for a
certain range of $\Phi$, holographic phase transitions are
obtained in the same dimensions as neutral black holes
\cite{DehghaniHendiMann2019} in massive gravity. Here, it
	should be emphasized that the critical behavior of charged
TBHs in massive gravity at the critical point are
indistinguishable if the effective topological factor
$k_{\rm{eff}}^{\rm{(C)}} \equiv [k + {m^2}c_0^2{c_2}]$ in the
canonical ensemble or $k_{\rm{eff}}^{\rm{(GC)}}$ in the grand
canonical ensemble have the same value while keeping other
parameters fixed.

Motivated by the fact that some characteristic features of universality class of phase transitions such as the critical exponents or universal ratio may depend on the ensemble or the spacetime dimensions, we discussed the universal ratio of critical phenomena at the critical point and also their critical exponents in both the ensembles. In the canonical ensemble up to two interaction potentials $O({\cal U}_2)$ (or equivalently $O(c_2)$), it is found that the universal ratio belongs to the universality class presented in Eq. (\ref{ratio-can-shifted}) which only depends on the spacetime dimensions whenever is written down in terms of the shifted Hawking temperature. In $d=4$, one arrives at $3/8$ for this ratio, exactly the same as vdW fluid. This result is the same as Einstein gravity, but holds for all types of massive gravity's TBHs in the same manner, in contrast to Einstein gravity in which only spherical black holes admits criticality. In the grand canonical ensemble up to three interaction potentials $O({\cal U}_3)$ (or equivalently $O(c_3)$), the critical ratio belongs to the another universality class (\ref{ratio-grand-shifted}) whenever is written down in terms of the shifted Hawking temperature. Interestingly for this case, the universal ratio is constant for all types of TBHs, i.e., it is independent of spacetime dimensions or any other parameter. Note that, according to Eqs. (\ref{critical temp-can}) and (\ref{critical temp-GC}), the shifted Hawking temperature at critical point is always positive. On the other hand, the universal ratio in both ensembles is a function of massive gravity's parameters ($m_g$ and $c_i$) and spacetime dimensions ($d$) whenever it is written down in terms of the standard Hawking temperature (see Eqs. (\ref{ratio-can}) and (\ref{ratio-grand})). So, the universal ratio depends on the specific choice of ensemble. These results are summarized in table (\ref{tab:universal-ratios}) for convenience.

\begin{table}[]
	\caption{Summary of universal ratios in the canonical and grand canonical ensembles. For the canonical ensemble, the results are derived up to two interaction potentials, $O({\cal U}_2)$. But in the case of grand canonical ensemble, the results are up to $O({\cal U}_3)$.}
	\label{tab:universal-ratios}
	\begin{tabular}{cccll}
		\cline{2-3}
		\multicolumn{1}{c|}{\textbf{}} & \multicolumn{1}{c|}{\textbf{\begin{tabular}[c]{@{}c@{}}Universal ratio in terms of\\ the shifted Hawking temperature\end{tabular}}} & \multicolumn{1}{c|}{\textbf{\begin{tabular}[c]{@{}c@{}}Universal ratio in terms of\\ the standard Hawking temperature\end{tabular}}} &  &  \\ \cline{1-3}
		\multicolumn{1}{|c|}{\textbf{Canonical ensemble}} & \multicolumn{1}{c|}{\begin{tabular}[c]{@{}c@{}} \large{$\frac{{{P_C}{r_c}}}{{{{\tilde T}_C}}} = \frac{{2d - 5}}{{16}}\Longleftrightarrow \frac{{{P_C}{v_c}}}{{{{\tilde T}_C}}} = \frac{{2d - 5}}{{4{d_2}}}$}\end{tabular}} & \multicolumn{1}{c|}{\begin{tabular}[c]{@{}c@{}}\,\\ \large{$\frac{{{P_C}{v_c}}}{{{T_C}}} = \frac{{(2d - 5)d_3^2 k_{\rm{eff}}^{\rm{(C)}}}}{{{d_2}\left( {4d_3^2 k_{\rm{eff}}^{\rm{(C)}}+ (2d - 5)m_g^2{c_0}{c_1}{r_c}} \right)}}$} \\ \,\end{tabular}} &  &  \\ \cline{1-3}
		\multicolumn{1}{|c|}{\textbf{\begin{tabular}[c]{@{}c@{}} Grand canonical\\ ensemble\end{tabular}}} & \multicolumn{1}{c|}{\begin{tabular}[c]{@{}c@{}} \,\\ \large{$\frac{{{P_C}{r_c}}}{{{{\tilde T}_C}}} = \frac{{d - 2}}{{12}} \Longleftrightarrow \frac{{{P_C}{v_c}}}{{{{\tilde T}_C}}} = \frac{1}{3}$} \\ \, \end{tabular}} & \multicolumn{1}{c|}{\begin{tabular}[c]{@{}c@{}} \,\\ \large{$\frac{{{P_C}{v_c}}}{{{T_C}}} = \frac{{{d_3}{d_4}c_0^3{c_3}}}{{3{d_3}{d_4}c_0^3{c_3} - {c_0}{c_1}r_c^2}}$} \\ \, \end{tabular}} &  &  \\ \cline{1-3}
		\multicolumn{1}{l}{} & \multicolumn{1}{l}{} & \multicolumn{1}{l}{} &  & 
	\end{tabular}
\end{table}

Furthermore, considering all the higher order self-interaction potentials of massive gravitons in arbitrary dimensions, we examined the associated critical exponents in the grand canonical ensemble and proved that they match to those of charged TBHs in the canonical ensemble (i.e., $\alpha=0$, $\beta=1/2$, $\gamma=1$	and $\delta=3$, exactly the same as vdW fluids).  So, all kinds of TBHs in massive gravity have the same critical exponents in arbitrary dimensions which indicates the universality class independent of the spacetime dimensions and also the ensemble one is dealing with.

In conclusion, the nature of TBH phase transitions depends on the ensemble and also spacetime dimensions. The RPT phenomenon only appeared in the grand canonical ensemble, while the rest of critical phenomena (vdW, triple point and vdW type) appear in both the ensembles, but they commence to show up in different dimensions (summarized in table \ref{tab:ensembles}). The universal ratio is also depends on the ensemble and spacetime dimensions (summarized in table \ref{tab:universal-ratios}). However, the critical exponents, which are the same as the exponents of vdW fluid, depend on neither spacetime dimensions nor ensemble.

\begin{acknowledgements}
We would like to thank the anonymous referees for their valuable comments which improved
the quality of this manuscript significantly. We wish to thank Shiraz University Research
Council. AD would like to thank Soodeh Zarepour for providing Mathematica programming
codes. The work of SHH has been supported financially by the Research Institute for
Astronomy and Astrophysics of Maragha, Iran.
\end{acknowledgements}


\begin{thebibliography}{}

    \bibitem{HawkingPage1983}S. Hawking, D.N. Page, Commun. Math. Phys. \textbf{87}, 577 (1983)

    \bibitem{EmparanChamblin1999a}A. Chamblin, R. Emparan, C.V. Johnson, R.C. Myers, Phys. Rev. D \textbf{60}, 064018 (1999)

    \bibitem{EmparanChamblin1999b}A. Chamblin, R. Emparan, C.V. Johnson, R.C. Myers, Phys. Rev. D \textbf{60}, 104026 (1999)

    \bibitem{Caldarelli2000KerrNewmanAdS}M.M. Caldarelli, G. Cognola, D. Klemm, Class. Quant. Grav. \textbf{17}, 399 (2000)

    \bibitem{KubiznakMann2012}D. Kubiznak, R.B. Mann, JHEP \textbf{07}, 033 (2012)

    \bibitem{Altamirano2012Reentrant}N. Altamirano, D. Kubiznak, R.B. Mann, Phys. Rev. D. \textbf{88}, 101502 (2013)

    \bibitem{Wei2016KerrNewmanAdS}P. Cheng, S-W Wei, Y-X Liu, Phys. Rev. D. \textbf{94}, 024025 (2016)

    \bibitem{NarayananKumar1994} T. Narayanan, A. Kumar, Phys. Rep. \textbf{249}, 135 (1994)

    \bibitem{Altamirano2014TriplePoint}N. Altamirano, D. Kubiznak, R.B. Mann, Z. Sherkatghanad, Class. Quant. Grav. \textbf{31}, 042001 (2014)

    \bibitem{Fernando(2006)BI-AdS}S. Fernando, Phys. Rev. D. \textbf{74}, 104032 (2006)

    \bibitem{Gunasekaran2012Mann}S. Gunasekaran, R.B. Mann, D. Kubiznak, JHEP \textbf{11}, 110 (2012)

    \bibitem{Vahidinia2013}S.H. Hendi, M.H. Vahidinia, Phys. Rev. D \textbf{88}, 084045 (2013)

    \bibitem{Cai2013GaussBonnet}R.G. Cai, L.M. Cao, L. Li, R.Q. Yang, JHEP \textbf{09}, 005 (2013)

    \bibitem{Zou2014GB-PV-GrandCan} D. Zou, Y. Liu, B. Wang, Phys. Rev. D \textbf{90}, 044063 (2014)

    \bibitem{PV2014Lovelock} H. Xu, W. Xu, L. Zhao, Eur. Phys. J. C \textbf{74}, 3074 (2014)

    \bibitem{PV2014LovelockBI-Mo} J.X. Mo, W.B. Liu, Eur. Phys. J. C \textbf{74}, 2836 (2014)

    \bibitem{PV2015Lovelock-nonExtended}J.X. Mo, W.B. Liu, Eur. Phys. J. C \textbf{75}, 211 (2015)

    \bibitem{PV2015LovelockBI-Belhaj}A. Belhaj, M. Chabab, H. EL Moumni, K. Masmar, M. B. Sedra, Int. J. Geom. Meth. Mod. Phys. \textbf{12}, 1550115 (2015)

    \bibitem{Frassino2014}A.M. Frassino, D. Kubiznak, R.B. Mann, F. Simovic, JHEP \textbf{09}, 080 (2014)

    \bibitem{HennigarBrennaMann2015}R.A. Hennigar, W.G. Brenna, R.B. Mann, JHEP \textbf{07}, 077 (2015)

    \bibitem{LovelockRainbow2017}S.H. Hendi, A. Dehghani, Mir Faizal, Nucl. Phys. B \textbf{914}, 117 (2017)

    \bibitem{Hennigar2017BlackBranesPV} R.A. Hennigar, JHEP \textbf{09}, 082 (2017)

    \bibitem{LovelockHairyBHs2017} R. A. Hennigar, E. Tjoa, R. B. Mann, JHEP \textbf{02}, 070 (2017)

    \bibitem{Mir2019Mann} M. Mir, R. A. Hennigar, J. Ahmed, R. B. Mann, JHEP \textbf{08}, 068 (2019)

    \bibitem{dRG2010} C. de Rham, G. Gabadadze, Phys. Rev. D \textbf{82}, 044020 (2010)

    \bibitem{dRGT2011} C. de Rham, G. Gabadadze, A.J. Tolley, Phys. Rev. Lett. \textbf{106}, 231101 (2011)

    \bibitem{Hendi2017Mann-PRD} S.H. Hendi, , R.B. Mann, S. Panahiyan, B. Eslam Panah, Phys. Rev. D. \textbf{95}, 021501 (2017)

    \bibitem{DehghaniHendiMann2019} A. Dehghani, S.H. Hendi, R.B. Mann, Phys. Rev. D \textbf{101}, 084026 (2020)

    \bibitem{Cai2015massive}R.G. Cai, Y.P. Hu, Q.Y. Pan, Y.L. Zhang, Phys. Rev. D \textbf{91}, 024032 (2015)

    \bibitem{PVmassive2015PRD}J. Xu, L.-M. Cao, Y.-P. Hu, Phys. Rev. D \textbf{91}, 124033 (2015)

    \bibitem{Reentrant-dRGTmassive-2017}D. Zou, R. Yue, M. Zhang, Eur. Phys. J. C \textbf{77}, 256 (2017)

    \bibitem{Triple-BI-massive-2017}M. Zhang, D. Zou, R. Yue, Adv. High Energy Phys. \textbf{2017}, Article ID 3819246 (2017)

    \bibitem{HendiDehghani2019}S.H. Hendi, A. Dehghani, Eur. Phys. J. C \textbf{79}, 227 (2019)

    \bibitem{Dehyadegari2019} A. Dehyadegari, B. R. Majhi, A. Sheykhi, A. Montakhab, Phys. Lett. B \textbf{791}, 30 (2019)

    \bibitem{Witten1998a}E. Witten, Adv. Theor. Math. Phys. \textbf{2}, 253 (1998)

    \bibitem{Witten1998b}W. Witten, Adv. Theor. Math. Phys. \textbf{2}, 505 (1998)

    \bibitem{Kastor2009CQG}D. Kastor, S. Ray, J. Traschen, Class. Quant. Grav. \textbf{26}, 195011 (2009)

    \bibitem{Dolan2011CQG1}B.P. Dolan, Class. Quant. Grav. \textbf{28}, 125020 (2011)

    \bibitem{Dolan2011CQG2}B.P. Dolan, Class. Quant. Grav. \textbf{28}, 235017 (2011)

    \bibitem{Dolan2011PRD}B.P. Dolan, Phys. Rev. D \textbf{84}, 127503 (2011)

    \bibitem{CveticKubiznak2011Gibbons-PRD}M. Cvetic, G. Gibbons, D. Kubiznak, C. Pope, Phys. Rev. D \textbf{84}, 024037 (2011)

    \bibitem{DolanKastorMann2013}B.P. Dolan, D. Kastor, D. Kubiznak, R.B. Mann, J. Traschen, Phys. Rev. D \textbf{87}, 104017 (2013)

    \bibitem{BlackHoleChemistry2017CQG}D. Kubiznak, R.B. Mann, M. Teo, Class. Quant. Grav. \textbf{34}, 063001 (2017)

    \bibitem{Galaxies2014}N. Altamirano, D. Kubiznak, R.B. Mann, Z. Sherkatghanad, Galaxies \textbf{2}, 89 (2014)

    \bibitem{HennigarMann2017lambdaLine}R.A. Hennigar, R.B. Mann, Phys. Rev. Lett. \textbf{118}, 021301 (2017)

    \bibitem{HassanRosen2012PRL} S.F. Hassan, R.A. Rosen, Phys. Rev. Lett. \textbf{108}, 041101 (2012)

    \bibitem{LIGO2017}LIGO scientific collaboration, Phys. Rev. Lett. \textbf{118}, 221101 (2017)

    \bibitem{deRham2014Review}C. de Rham, Living Rev. Rel. \textbf{17}, 7 (2014)

    \bibitem{Schmidt-May2016DarkMatter}E. Babichev, L. Marzola, M. Raidal, A. Schmidt-May, F. Urban, H. Veermae, M. von Strauss, J. Cosmol. Astropart. Phys. \textbf{09}, 016 (2016)

    \bibitem{MassiveCosmology2013}Y. Akrami, T.S. Koivisto, M. Sandstad, JHEP \textbf{03}, 99 (2013)

    \bibitem{MassiveCosmology2015}Y. Akrami, S.F. Hassan, F. Knnig, A. Schmidt-May, A.R. Solomon, Phys. Lett. B \textbf{748}, 37 (2015)

    \bibitem{MGinString2018}C. Bachas , I. Lavdas, JHEP \textbf{11}, 003 (2018)

    \bibitem{Hinterbichler2012Review} K. Hinterbichler, Rev. Mod. Phys. \textbf{84}, 671 (2012)

    \bibitem{HassanRosen2012JHEP} S.F. Hassan, R.A. Rosen, A. Schmidt-May, JHEP \textbf{02}, 026 (2012)

    \bibitem{TQDo2016a}T.Q. Do, Phys. Rev. D \textbf{93}, 104003 (2016)

    \bibitem{TQDo2016b}T.Q. Do, Phys. Rev. D \textbf{94} 044022 (2016)

    \bibitem{Vegh2013}D. Vegh, [arXiv:1301.0537]

    \bibitem{Hendi2016JHEP} S.H. Hendi, S. Panahiyan, B. Eslam Panah, JHEP \textbf{01}, 129 (2016)

    \bibitem{Gabadadze2018Pirtskhalava} G. Gabadadze, D Pirtskhalava, Phys. Rev. D \textbf{97}, 124045 (2018)

    \bibitem{Alberte2013} L. Alberte, A. Khmelnitsky, Phys. Rev. D \textbf{88}, 064053 (2013)

    \bibitem{Alberte2015}L. Alberte, A. Khmelnitsky, Phys. Rev. D \textbf{91}, 046006 (2015)

    \bibitem{BlakeTong2013} M. Blake, D. Tong, Phys. Rev. D \textbf{88}, 106004 (2013)

    \bibitem{Davison2013} R.A. Davison, Phys. Rev. D \textbf{88}, 086003 (2013)

    \bibitem{HHH2008PRL} S.A. Hartnoll, C.P. Herzog, G.T. Horowitz, Phys. Rev. Lett. \textbf{101}, 031601 (2008)

    \bibitem{HHH2008JHEP} S.A. Hartnoll, C.P. Herzog, G.T. Horowitz, JHEP \textbf{12}, 015 (2008)

    \bibitem{Gregory2009GBsuperconductor} R. Gregory, S. Kanno, J. Soda, JHEP \textbf{10}, 010 (2009)
    
    \bibitem{Gregory2010GBsuperconductor} L. Barclay, R. Gregory, S. Kanno, P. Sutcliffe, JHEP \textbf{12}, 029 (2010)

    \bibitem{Alberte2016SolidHolography} L. Alberte, M. Baggioli, A. Khmelnitsky, O. Pujolas, JHEP \textbf{02}, 114 (2016).

    \bibitem{Alberte2018HolographicPhonons} L. Alberte,M. Ammon, A. Jiménez-Alba, M. Baggioli, Oriol Pujolas, Phys. Rev. Lett. \textbf{120}, 171602 (2018).

    \bibitem{Alberte2018BHelasticity} L. Alberte, M. Ammon, M. Baggioli, A. Jiménez, O. Pujolàs, JHEP \textbf{01}, 129 (2018).
    
    \bibitem{Beekman2017PhysRep}A.J. Beekman, J. Nissinen, K. Wu, K. Liu, R-J Slager, Z. Nussinov, V. Cvetkovic, J. Zaanen, Physics Reports \textbf{683}, 1 (2017)
    
    \bibitem{Beekman2017PRB}A. J. Beekman, J. Nissinen, K. Wu, and J. Zaanen, Phys. Rev. B \textbf{96}, 165115 (2017)
   
    
    \bibitem{GhostFreeSingular2016} H. Zhang, XZ Li, Phys. Rev. D \textbf{93}, 124039 (2016)
    
    \bibitem{Koyama2011PRL} K. Koyama, G. Niz, G. Tasinato, Phys. Rev. Lett. \textbf{107}, 131101 (2011) 
    
    \bibitem{Nieuwenhuizen2011PRD} T.M. Nieuwenhuizen, Phys. Rev. D \textbf{84}, 024038 (2011)
    
    \bibitem{GruzinovMirbabayi2011PRD} A. Gruzinov, M. Mirbabayi, Phys. Rev. D \textbf{84}, 124019 (2011)
    
    \bibitem{deRham2011BHs} L. Berezhiani, G. Chkareuli, C. de Rham, G. Gabadadze, A.J. Tolley, Phys.Rev.D \textbf{85}, 044024 (2011)
    
    \bibitem{ChargedBHs2013PRD} Y-F Cai, D. A. Easson, C. Gao, E. N. Saridakis, Phys. Rev. D \textbf{87}, 064001 (2013)
     
    \bibitem{Babichev2014JHEP} E. Babichev, A. Fabbri, JHEP \textbf{07}, 016 (2014)
    
    \bibitem{Babichev2015CQG} E. Babichev, R. Brito, Class. Quant. Grav. \textbf{32}, 154001 (2015)

     \bibitem{York1990} H. W. Braden, J. D. Brown, B. F. Whiting, J. W. York Jr., Phys. Rev. D \textbf{42}, 3376 (1990)

     \bibitem{Nastase2015Book} H. Nastase, \textit{Introduction to the AdS/CFT Correspondence} (Cambridge University Press, 2015)

    \bibitem{Katsuragawa2015}T. Katsuragawa, S. Nojiri, Phys. Rev. D \textbf{91}, 084001 (2015)

    \bibitem{Klauber2013} R. D. Klauber, \textit{Student friendly quantum field theory: basic principles and quantum electrodynamics} (Sandtrove Press, Fairfield, Iowa, 2013)

    \bibitem{Zee2010QFT} A. Zee, \textit{Quantum Field Theory in a Nutshell} (Princeton University
    Press, Princeton, 2010)

    \bibitem{Natsuume2015AdS/CFT} M. Natsuume, \textit{AdS/CFT Duality User Guide} (Springer, Germany, 2015)

    \bibitem{Erdmenger2015} M. Ammon, J. Erdmenger, \textit{Gauge/gravity duality: Foundations and applications} (Cambridge University Press, Cambridge, 2015)

    \bibitem{BardeenCarterHawking1973}J.M. Bardeen, B. Carter, S.W. Hawking, Comm. Math. Phys. \textbf{31}, 161 (1973)

    \bibitem{Kastor2010LovelockSmarr}D. Kastor, S. Ray, J. Traschen, Class. Quant. Grav. \textbf{27}, 235014 (2010)

    \bibitem{Mathematica} Wolfram Research Inc., Mathematica, Version 11.0, Champaign, IL USA (2019)

    \bibitem{Wei2014}S.W. Wei, Y.X. Liu, Phys. Rev. D \textbf{90}, 044057 (2014)

    \bibitem{Huang2001Book} K. Huang, \textit{Statistical Mechanics} (CRC Press, 1987)

    \bibitem{Zinn-Justin2007} J. Zinn-Justin, \textit{Phase transitions and renormalization group} (Oxford University Press, New York, 2007)






\end{thebibliography}
\end{document}